# Gratings:
# Theory and Numeric Applications
## Second Revisited Edition

Tryfon Antonakakis
Fadi Baïda
Abderrahmane Belkhir
Kirill Cherednichenko
Shane Cooper
Richard Craster
Guillaume Demesy
John DeSanto
Gérard Granet

Boris Gralak
Leonid Goray
Sébastien Guenneau
Lifeng Li
Daniel Maystre
André Nicolet
Gunther Schmidt
Brian Stout
Fréderic Zolla

Evgeny Popov, Editor

Institut Fresnel, Université d'Aix-Marseille, Marseille, France
Institut FEMTO-ST, Université de Franche-Comté, Besançon, France
Institut Pascal, Université Blaise Pascal, Clermont-Ferrand, France
Colorado School of Mines, Golden, USA
CERN, Geneva, Switzerland
Imperial College London, UK
Cardiff University, Cardiff, UK
Université Mouloud Mammeri, Tizi-Ouzou, Algeria
Saint Petersburg Academic University, Saint Petersburg, Russian Federation
Institute for Analytical Instrumentation, Saint Petersburg, Russian Federation
Weierstrass Institute of Applied Analysis and Stochastics, Berlin, Germany
Tsinghua University, Beijing, China

Chapter 4:

Integral Method for Gratings

Daniel Maystre and Evgeny Popov

**Table of Contents:**

# Chapter 4

# Integral Method for Gratings

Daniel Maystre and Evgeny Popov

*Aix-Marseille Université, CNRS, Centrale Marseille, Institut Fresnel UMR 7249,
Campus de Saint Jerome,13013 Marseille, France*
*Daniel.maystre@fresnel.fr*

## 4.1. Introduction

Integral methods for scattering problems represent a class of mathematical methods based on integral equations. In this chapter, all the integral equations are deduced from boundary value problems of scattering and are classified as Fredholm integral equations. They can be written in the form:

$$c\,u(\vec{r}) = v(\vec{r}) + \int_{\mathbb{A}} W(\vec{r},\vec{r}')u(\vec{r}')d\vec{r}', \qquad (4.1)$$

in which $\mathbb{A}$ may represent for example the surface of a three-dimension diffracting object or the cross-section boundary of a two-dimension (cylindrical) object, v and u are continuous functions in $\mathbb{A}$. The kernel W of the equation is also continuous in $\mathbb{A}$, and $c = 0$ or 1 according to whether the equation is of the first or second kind. The mathematical problem is to determine u if v and W are known [1]. This kind of method is widely employed in many domains of physics [1,2], where it is usual to extend it to cases in which u, v, and W are only peasewise continuous or can even be singular. A typical example of use in electromagnetism can be found by applying the second theorem of Green for expressing the field in a given volume in terms of its values and of the values of its normal derivative on the surrounding surface. In that case, the kernels of the integral equations include combinations of Green's functions and of their normal derivatives on the boundaries of the scattering objects.

The theory of integral equations can be described in a rigorous, elegant and concise way using distribution theory [3-5] that extends derivatives and other differential operators to discontinuous functions (a famous example is the so-called Dirac function that, for the mathematician, should not be called function). The interested reader can find a detailed presentation of rigorous use of the distribution theory in the electromagnetic theory of gratings in [6,7].

The first application of the integral method in grating theory was proposed almost simultaneously for the case of perfectly conducting gratings by Petit and Cadilhac [8], Wirgin [9], and Uretski [10]. The first numerical implementation was reported by Petit [11,12] for TE polarization (called also P, or $E_{//}$, or s polarization). The first extension of this integral equation to the other polarization (TM or S or p) led to a non-integrable kernel. This problem was solved by Pavageau, who proposed for both polarizations new integral equations having continuous and bounded kernels [13].

Soon after, Wirgin [14], Neureuther and Zaki [15], and Van den Berg [16] gave formulations of the integral method applied to gratings made of metals with finite conductivity or dielectrics. The approach was based on the resolution of two coupled integral equations containing two unknown functions. Problems of limited memory storage and time-computation on computers in the late '60s restricted the numerical implementation of this theory to dielectric gratings only, for which very rare numerical results were published. This restriction was not considered as dramatic by grating specialists at that time. Indeed, it was generally considered that in the visible and infrared regions in which the reflectivity of usual metals exceeds 90%, the model of perfectly conducting grating is accurate. However, development of space optics and astronomy at that time required a precise treatement of the problem of metallic gratings used in the ultraviolet, a domain where the metal reflectivity starts to drop down as approaching electron plasma frequencies. This need required a new approach proposed by Maystre in 1972 [17] by using a single integral equation for a single unknown function. Solving the difficulties in the summation of the series in the kernels and in their integration, the integral method in this formulation was the first one to result in a computer code that was able to correctly model diffraction gratings behaviour over the entire spectrum for almost all commercial gratings profiles [18]. One of the most important conclusions for the practical applications and grating manufacturers was the definite demonstration of the inadequacy of the model of perfectly conducting grating in the near-infrared, visible and ultraviolet regions [19].

However, the method was unable to treat some kinds of gratings, for example gratings having large periods and steep facets (echelle gratings, for example) or gratings with profile that cannot be represented by Fourier series (rod gratings, cavity gratings, etc.). A further development of this approach was proposed by Maystre [20]. It was numerically implemented in the early '90s in the code 'Grating 2000' by the author, which is used in many academic and industrial centers in the world. Other development was required for other exotic cases that started to find applications and thus required theoretical support for modeling. This development covered conical mountings and specially gratings with dielectric multilayer coatings [21,22], buried gratings and bimetallic gratings [23-26].

It must be emphasized that in this chapter, the authors use, without complete demonstrations, some analytic properties of gratings demonstrated in chapter 2. Thus, it is recommended to read this chapter before the present one.

First, we will deal with the most frequent problem: the bare metallic or dielectric grating. Then, extensions will be given to other kinds of gratings like perfectly conducting gratings, dielectric coated gratings or gratings in conical diffraction.

**4.2. The integral method applied to a bare, metallic or dielectric grating.**

*4.2.1. The physical model*

The grating surface S of period d separates a region $V^+$ with real relative electric permittivity and magnetic permeability $\varepsilon^+$ and $\mu^+$ respectively and a region $V^-$ with real or complex relative electric permittivity and magnetic permeability $\varepsilon^-$ and $\mu^-$ (figure 4.1). The indices $n^+$ and $n^-$ of these media are given by $n^+ = \sqrt{\varepsilon^+\mu^+}$ and $n^- = \sqrt{\varepsilon^-\mu^-}$. We consider the classical diffraction case with incident wavevector $\vec{k}^i$ lying in the xz plane, i.e. the plane perpendicular to the grooves. The incidence angle $\theta^i$ is measured in the counterclockwise sense from the z axis and $\lambda = \frac{2\pi}{k}$ denotes the wavelength of light in vacuum. The ordinate of the top of the profile is denoted by $z_0$ and unit normal $\vec{N}_S$ is oriented towards $V^+$. We denote by s the curvilinear abscissa on S, with origin being located at the origin of the Cartesian coordinates, and $s_d$ denoting the curvilinear abscissa of the point of S of abscissa $x = d$.

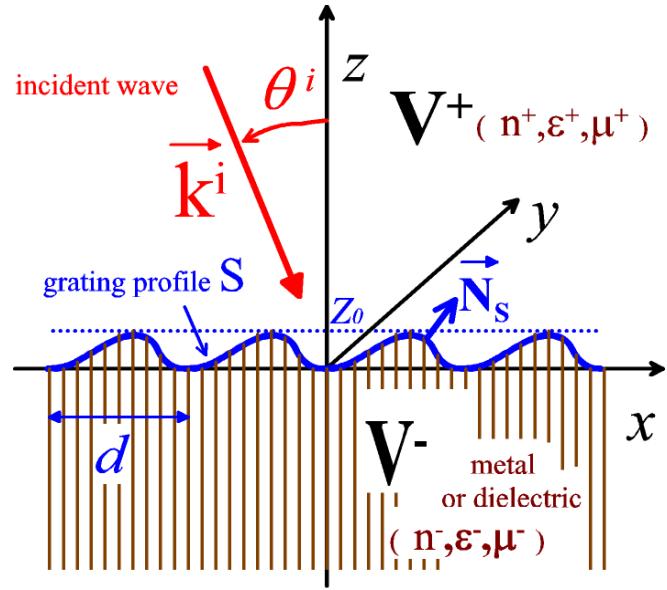

*Figure 4.1. Notations*

In this chapter, we use the complex notation with a time-dependence in $\exp(-i\omega t)$. Let F denote the y-component of the electromagnetic field. In TE polarization, it stays for the y-component $E_y$ of the electric field, and in TM case for the y-component $H_y$ of the magnetic field.

### 4.2.2. The boundary value problem.

It is shown in chapter 2 that in that case of classical diffraction, the total field $F^T = \begin{cases} F^{T+} & \text{in } V^+, \\ F^{T-} & \text{in } V^- \end{cases}$ is invariant along the y axis and that it is pseudo-periodic in x:

$$F^T(x+d,z) = F^T(x,z)\exp(i\alpha_0 d), \tag{4.2}$$

with:

$$\alpha_0 = k n^+ \sin\theta^i, \qquad k = \frac{2\pi}{\lambda}. \tag{4.3}$$

Moreover, the scattered field is defined by:

$$F = \begin{cases} F^+ = F^{T+} - F^i & \text{in } V^+, \\ F^- = F^{T-} & \text{in } V^-, \end{cases} \tag{4.4}$$

with the incident field $F^i$ in $V^+$ being given by:

$$F^i(x,z) = e^{i\alpha_0 x - ikn^+ \cos\theta^i z}. \tag{4.5}$$

The interest of the notion of scattered field is that it satisfies a radiation condition (also called Sommerfeld condition, or outgoing wave condition, see chapter 2) for $z \to \pm\infty$. The radiation condition states that the scattered field at infinity must remain bounded and must propagate upward in $V^+$ and dawnward in $V^-$. The scattered field also satisfies Helmholtz equations:

$$\nabla^2 F^\pm + k^2 \left(n^\pm\right)^2 F^\pm = 0 \quad \text{in } V^\pm. \tag{4.6}$$

The invariance along the y axis allows one to reduce the scattering problem to a two-dimension problem while the pseudo-periodicity restricts the study of the field to a single period of the grating. Consequently, it can be considered that $V^\pm$ are no more volumes but surfaces extending on a single period of the grating.

In order to periodize the scattered field, we introduce the function $U = \begin{cases} U^+ \text{ in } V^+, \\ U^- \text{ in } V^-, \end{cases}$:

$$U(x,z) = e^{-i\alpha_0 x} F(x,z). \tag{4.7}$$

We denote by $\psi^\pm(s)$ the limit values of $U^\pm$ on S and $\phi^\pm(s)$ the values of $e^{-i\alpha_0 x} \dfrac{dF^\pm}{dN_S}$, with $\dfrac{dF^\pm}{dN_S}$ being the normal derivative of $F^\pm$ on S.

The Helmholtz equations and the radiation condition are not sufficient to define the boundary value problem satisfied by the scattered field F. A third kind of condition must be added: the boundary conditions of the electromagnetic field components across S. The tangential components of the electric and magnetic fields are continuous across this interface, as far as the permittivities and permeabilities of the two media take finite values. For both polarizations, this property yields:

$$\psi^+(s) + \psi^i(s) = \psi^-(s), \tag{4.8}$$

with $\psi_i$ being the value of the periodized incident field, obtained from equation (4.5):

$$\psi^i = F^i \left[x(s), z(s)\right] \exp(-i\alpha_0 x(s)) = \exp\left[-ikn^+ \cos\theta^i z(s)\right]. \tag{4.9}$$

Using Maxwell equations, the continuity of the tangential component of the magnetic field (for TE polarization) and that of the electric field (for TM polarization) leads to the following relation:

$$q^+ \left[\phi^+(s) + \phi^i(s)\right] = q^- \phi^-(s), \tag{4.10}$$

with

$$\phi^i = \exp(-i\alpha_0 x) \frac{\partial F^i}{\partial N_S} =$$
$$= -ikn^+ \left(\frac{dz(s)}{ds}\sin\theta^i + \frac{dx(s)}{ds}\cos\theta^i\right)\exp\left(-ikn^+ \cos\theta^i z(s)\right), \tag{4.11}$$

and

$$q^{\pm} = \begin{cases} \dfrac{1}{\mu^{\pm}}, & \text{for TE polarization,} \\ \dfrac{1}{\varepsilon^{\pm}}, & \text{for TM polarization.} \end{cases} \qquad (4.12)$$

### *4.2.3. Integral equation*

The theoretical basis of the integral method lies on a general property of the electromagnetic field: the field inside a given surface of the xz plane can be expressed from the values of the field and of its normal derivatives on the curve surrounding the surface, according to the second Green's theorem. The value of $U^{\pm}$ at a point of $V^{\pm}$ of coordinates x and z be deduced from its values on S using equation (4.139) of appendix 4.A:

$$U^{\pm}(x,z) = \pm \int_{s'=0}^{S_d} \left[ \mathcal{G}^{\pm}(x,z,s')\phi^{\pm}(s') + \mathcal{N}^{\pm}(x,z,s')\psi^{\pm}(s') \right] ds', \qquad (4.13)$$

with

$$\mathcal{G}^{\pm}(x,z,s') = \frac{1}{2id} \sum_{m=-\infty}^{\infty} \frac{1}{\gamma_m^{\pm}} \exp\left\{ imK\left[x-x'(s')\right] + i\gamma_m^{\pm}\left|z-z'(s')\right| \right\}, \qquad (4.14)$$

$$\mathcal{N}^{\pm}(x,z,s') = \frac{1}{2d} \sum_{m=-\infty}^{\infty} \left\{ \frac{dx(s')}{ds'} \operatorname{sgn}\left[z-z'(s')\right] - \frac{\alpha_m}{\gamma_m^{\pm}} \frac{dz'(s')}{ds'} \right\} \times \\ \times \exp\left\{ imK\left[x-x'(s')\right] + i\gamma_m^{\pm}\left|z-z'(s')\right| \right\}, \qquad (4.15)$$

where:

$$\alpha_m = \alpha_0 + mK, \qquad (4.16)$$

$$K = \frac{2\pi}{d}, \qquad (4.17)$$

$$\gamma_m^{\pm} = \sqrt{\left(kn^{\pm}\right)^2 - \alpha_m^2}, \qquad (4.18)$$

with s′ being the curvilinear abscissa on a point of S with coordinates x′(s′) and z′(s′).

According to section 4.A.4, the values of $\psi^{\pm}(s')$ and $\phi^{\pm}(s')$ are linked by a relation of compatibility. Using eqs. (4.146), (4.147) and (4.148) we obtain:

$$\int_{s'=0}^{S_d} \left[ \mathcal{G}^{+}(s,s')\tilde{\phi}^{+}(s') + \mathcal{N}^{+}(s,s')\tilde{\psi}^{+}(s') \right] ds' - \frac{\tilde{\psi}^{+}(s')}{2} = 0, \qquad (4.19)$$

$$\int_{s'=0}^{S_d} \left[ \mathcal{G}^{-}(s,s')\tilde{\phi}^{-}(s') + \mathcal{N}^{-}(s,s')\tilde{\psi}^{-}(s') \right] ds' + \frac{\tilde{\psi}^{-}(s')}{2} = 0, \qquad (4.20)$$

with:

$$\mathcal{G}^{\pm}(s,s') = \frac{1}{2id} \sum_{m=-\infty}^{\infty} \frac{1}{\gamma_m^{\pm}} \exp\left[imK(x(s)-x'(s')) + i\gamma_m^{\pm}|z(s)-z'(s')|\right], \qquad (4.21)$$

$$\mathcal{N}^{\pm}(s,s') = \frac{1}{2d} \sum_{m=-\infty}^{\infty} \left[\frac{dx'}{ds'}\text{sgn}(z(s)-z'(s')) - \frac{\alpha_m}{\gamma_m^{\pm}}\frac{dz'}{ds'}\right] \times \qquad (4.22)$$

$$\times \exp\left[imK(x(s)-x'(s')) + i\gamma_m^{\pm}|z(s)-z'(s')|\right].$$

Introducing in eq. (4.20) the values of $\psi^-(s) = \psi^+(s) + \psi_i(s)$ and $\phi^-(s) = \frac{q^+}{q^-}\left[\phi^+(s) + \phi_i(s)\right]$ given by the continuity conditions on the grating profile (eqs (4.8), (4.9), (4.10), (4.11) and (4.12)) yields a second integral equations with two unknown functions $\psi^+$ and $\phi^+$:

$$\int_{s'=0}^{S_d} \left\{\frac{q^+}{q^-}\mathcal{G}^-(s,s')\left[\phi^+(s') + \phi^i(s')\right] + \mathcal{N}^-(s,s')\left[\psi^+(s') + \psi^i(s')\right]\right\}ds' \qquad (4.23)$$

$$+ \frac{\psi^+(s') + \psi^i(s')}{2} = 0.$$

Eqs (4.19) and (4.23) constitute a system of two integral equations with two unknown functions, which can be solved on a computer. The amplitudes $r_m$ and $t_m$ of the plane waves reflected and transmitted by the grating can be deduced from the solution of the integral equation using eqs. (4.184) and (4.185) of appendix 4.A:

$$r_m = \frac{1}{2d}\int_{s=0}^{S_d} e^{-imKx(s) - i\gamma_m^+ z(s)}\left[\frac{-i\phi^+(s)}{\gamma_m^+} + \left(\frac{dx(s)}{ds} - \frac{\alpha_m}{\gamma_m^+}\frac{dz(s)}{ds}\right)\psi^+(s)\right]ds, \qquad (4.24)$$

$$t_m = \frac{1}{2d}\int_{s=0}^{S_d} e^{-imKx(s) + i\gamma_m^- z(s)}\left[\frac{i\phi^-(s)}{\gamma_m^-} + \left(\frac{dx(s)}{ds} + \frac{\alpha_m}{\gamma_m^+}\frac{dz(s)}{ds}\right)\psi^-(s)\right]ds, \qquad (4.25)$$

with $z_0$ being the ordinate of the top of the grating profile. For non-evenescent reflected orders, diffraction efficiencies $\rho_m$ can be obtained using eq. (4.187):

$$\rho_m = \frac{\gamma_m^+}{\gamma_0^+}|r_m|^2. \qquad (4.26)$$

For gratings made of a lossless dielectric material in $V^-$, transmitted efficiencies can be defined as well:

$$\tau_m = \frac{q^-}{q^+}\frac{\gamma_m^+}{\gamma_0^+}|t_m|^2. \qquad (4.27)$$

In that case the energy balance (see chapter 2) can be expressed by:

$$\sum_{m \in P^+} \rho_m + \sum_{m \in P^-} \tau_m = 1, \qquad (4.28)$$

with $P^+$ and $P^-$ denoting respectively the set of non-evanescent reflected and transmitted orders. The numerical implementation of the integral equations will be described in section 6. In contrast with the two coupled equations obtained in this section, the integral equation obtained by Maystre is unique. This feature requires the definition of a single and well adapted unknown function. The mathematical definition of this function needs the use of tools of applied mathematics described in appendix A. Appendix B contains a mathematical description of this mathematical function and of the integral equation. Here, we give a heuristic description of this function for TE polarization. First, we replace the material in $V^-$ by the same material as in $V^+$, the entire space being thus homogeneous. It can be shown that it exists one (and only one) distribution of surface current density $\Phi$ parallel to the y axis, placed on S (this surface separates now two identical media), which generates in $V^+$ a field equal to the actual diffracted field in the physical problem. Intuitively, it is easy, from this surface current density, to express in an integral form the actual scattered field in $V^+$, this current distribution being nothing else than a set of current lines placed in a homogeneous medium. From the expression of the scattered field in $V^+$, simple mathematical calculations allow one to deduce the scattered field and its normal derivative above S, thus $\psi^+(s)$ and $\phi^+(s)$ from the unique unknown function $\Phi$.

Now, we abandon the field generated by the fictitious surface current density $\Phi$, except the integral expressions of $\psi^+(s)$ and $\phi^+(s)$ containing $\Phi$, and we come back to the actual physical grating problem. The continuity conditions for the tangential components of the field permit the calculation of $\psi^-(s)$ and $\phi^-(s)$ thus, using the second Green theorem, of the actual physical field below S. At that point it has been shown that the four unknown functions contained in the classical theory previously described in this section can be derived from a single one and that, in some way, there is a redundancy in the use of multiple unknown functions. It is easy to understand that this single unknown function can be calculated from a single integral equation. This equation can be obtained for example by writing the continuity on S of the integral expressions of the field in $V^+$ and $V^-$.

This method was the first one to show that, in contrast with the second Green theorem, it exists a formula that allows one to express the field inside a given domain from a single function defined on its boundary. This function is neither the field nor its normal derivative, but both can be deduced from it through simple integrals. These integrals automatically satisfy the compatibility condition.

### 4.3. The bare, perfectly conducting grating

Perfectly conducting gratings were historically the first gratings to be modeled using rigorous electromagnetic theories. They represent accurate models for metallic gratings working in far infrared and microwaves regions. The pioneering works appear in the '60s [8] and were followed by many papers. Various formulations of the integral method have been published. They differ either in the form of the integral equation ot in the numerical implementation. A review of this matter may be found in [27, 6, 28].

#### *4.3.1. Perfectly conducting gratings in TE polarization*

Two different approaches will be presented in this section. The first one, published by R. Petit and M. Cadilhac in [8], leads to a Fredholm integral equation of the fist kind with a singular kernel. A version leading to a Fredholm integral equation of the second kind with a non-singular kernel was proposed by Pavageau et al. [13] using the ideas of Maue [29].

The total field in $V^+$ is pseudo-periodic, it satisfies the Helmholtz equation:

$$\nabla^2 F^{T+} + k^2 \left(n^+\right)^2 F^{T+} = 0 \quad \text{in } V^+, \tag{4.29}$$

and a radiation condition at infinity. Thus we can apply the generalized compatibility condition of section 4.A.5:

$$\int_{s'=0}^{s_d} \left[\mathcal{G}^+(s,s')\Phi^+(s') + \mathcal{N}^+(s,s')\Psi^+(s')\right]ds' + \psi^i(s) = \frac{\Psi^+(s)}{2}, \tag{4.30}$$

with $\Psi^+(s')$ and $\Phi^+(s')$ denoting the limit of the total field on S and its normal derivative. The boundary condition on S is straightforward: the total electric field, which is parallel to the y axis thus tangential to the metal, vanishes on S. This property entails that $\Psi^+(s) = 0$ and thus, eq. (4.30) becomes, in operator notation:

$$\mathcal{G}^+ \Phi^+ = -\psi^i. \tag{4.31}$$

with $\mathcal{G}^+(s,s')$ given by eq. (4.147)

$$\mathcal{G}^+(s,s') = \frac{1}{2id} \sum_{m=-\infty}^{\infty} \frac{1}{\gamma_m^+} \exp\left[imK\left(x(s) - x'(s')\right) + i\gamma_m^+ \left|z(s) - z'(s')\right|\right]. \tag{4.32}$$

This is a Fredholm integral equation of the first kind, with a singular kernel.

The amplitudes of the reflected waves are deduced from eq. (4.186):

$$r_m = \frac{1}{2id\gamma_m^+} \int_{s=0}^{s_d} \exp\left[-imKx(s) - i\gamma_m^+ z(s)\right] \Phi^+(s) ds. \tag{4.33}$$

The efficiencies $\rho_m = \frac{\gamma_m^+}{\gamma_0^+} \left|r_m\right|^2$ satisfy the energy balance relation:

$$\sum_{P^+} \rho_m = 1, \tag{4.34}$$

with $P^+$ denoting the set of non-evanescent orders.

An integral equation of the second kind with a regular continuous kernel can be found using the same function $\Phi^+$. It is shown in section 4.A5 that the normal derivative $\frac{dF^+}{dN_S}$ can be calculated in that case (eq. (4.172)). This integral equation can be written either by writing that $\frac{dF^+}{dN_S} = \phi^+ e^{i\alpha_0 x}$ is equal to $\left(\Phi^+ - \phi^i\right) e^{i\alpha_0 x}$. The final equation is given by:

$$\frac{\Phi^+(s)}{2} = \phi^i(s) + \int_{s'=0}^{s_d} \mathcal{K}(s,s') e^{i\alpha_0 x} \Phi^+(s) ds', \tag{4.35}$$

with:

$$\mathcal{K}(s,s') = \frac{1}{2d} \sum_{m=-\infty,+\infty} \left[ \text{sgn}(z-z')\frac{dx}{ds} - \frac{\alpha_m}{\gamma_m^+}\frac{dz}{ds} \right] e^{imK(x-x')+i\gamma_m^+|z-z'|}. \tag{4.36}$$

### 4.3.2. Perfectly conducting gratings for TM polarization

Once again, the generalized compatibility condition is used (eq. (4.158):

$$\int_{s'=0}^{S_d} \left[ \mathcal{G}^+(s,s')\Phi^+(s') + \mathcal{N}^+(s,s')\Psi^+(s') \right] ds' + \psi^i(s) = \frac{\Psi^+(s)}{2}. \tag{4.37}$$

In that case too, the tangential component of the electric field vanishes on the profile. It is shown in chapter 2 from Maxwell equations that this condition entails that the normal derivative of the total field vanishes on S, thus:

$$\Phi^+ = 0, \tag{4.38}$$

so that

$$\int_{s'=0}^{S_d} \mathcal{N}^+(s,s')\Psi^+(s')ds' + \psi^i(s) = \frac{\Psi^+(s)}{2}, \tag{4.39}$$

with $\mathcal{N}^+(s,s')$ given by eq. (4.22) and (4.148):

$$\mathcal{N}^+(s,s') = \frac{1}{2d} \sum_{m=-\infty}^{\infty} \left[ \frac{dx'}{ds'}\text{sgn}(z(s)-z'(s')) - \frac{\alpha_m}{\gamma_m^\pm}\frac{dz'}{ds'} \right] \times$$
$$\times \exp\left[ imK(x(s)-x'(s')) + i\gamma_m^+|z(s)-z'(s')| \right]. \tag{4.40}$$

The Fredholm integral equation of the second kind with a regular continuous kernel is very close to that obtained for TE polarization (eq. (4.35)).
The amplitudes of the reflected waves are deduced from eq. (4.186):

$$r_m = \frac{1}{2d} \int_{s=0}^{S_d} \exp\left[ -imKx(s) - i\gamma_m^+ z(s) \right] \left( \frac{dx(s)}{ds} - \frac{\alpha_m}{\gamma_m^+}\frac{dz(s)}{ds} \right) \Psi(s)ds, \tag{4.41}$$

As for TE polarization, the efficiencies $\rho_m = \frac{\gamma_m^+}{\gamma_0^+}|r_m|^2$ satisfy the energy balance relation:

$$\sum_{P^+} \rho_m = 1. \tag{4.42}$$

### 4.4. Multiprofile gratings

The use of dielectric coatings has many applications even for diffraction gratings use. For example, metallic gratings are covered by a thin layer of dielectric material in order to avoid oxidation of the metal. Dielectric gratings can require an antireflection coating consisting of a thin layer or a stack of layers. Conversely, a stack of layers is used to increase the metal reflectivity or even to replace it, in order to reduce the absorption of light beams for high power laser applications.

Up to our knowledge, the first numerical results in the study of coated gratings was made by Van de Berg [16] for a single-layer perfectly conducting grating with the layer filling up the space between the grating surface and a plane surface. Although at that time the interest in such geometry remained mostly academic, further development of technology made it possible to fabricate such layer by dielectric coating and polishing. Another important application comes from the process of replication of dielectric gratings using an epoxy layer to transfer the replica to a plane surface of the substrate, or to have epoxy as grating layer itself.

After this initial work, two integral methods were proposed. The first one [20] is theoretically able to deal with an arbitrary multilayer grating without limitations concerning the shape of the profile or the conductivity of the layers. The second method [21] can deal with a multilayer grating without interpenetration of the profiles.Botten has solved the problem with a single-profile grating that has a stack of plane layers below and, eventually, above it [22, 23], by introducing a new form of Green's function, adapted to the multilayer system, which leeds to a single integral equation.

In what follows, we will at first describe the method for a single interface inside a stack, when the layer is relatively thin so that the upper and the lower interface interpenetrate. It is important to distinguish the two cases, with and without interpenetration, because in the latter case, it is possible to define a plane layer in between that does not cross the upper or the lower interface. This possibility enables one to use the plane-wave Rayleigh expansion of the electromagnetic field between the interfaces, whereas in the former case it is necessary to write and to solve a system of coupled integral equations that link the field components on the top and bottom of each layer.

### *4.4.1. Thin-layer gratings*

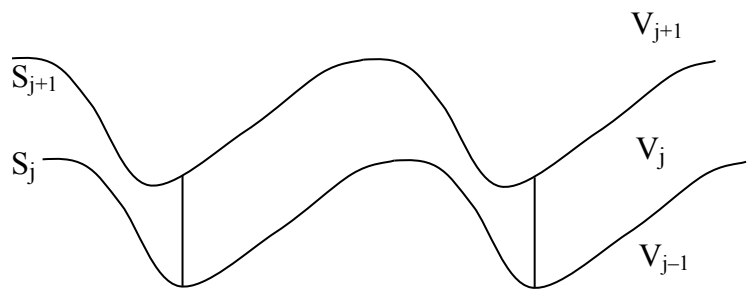

*Figure 4.2. Single layer inside a stack of a multilayer grating*

Let us consider the case of a multilayer grating having profiles that interpenetrate. In other words, it is impossible to introduce inside the layer a plane surface that does not cross one of the profiles. Then it is impssible to use the plane-wave expansion between the interfaces and we are led to solve integral equations that are coupled on the two interfaces of each layer.

We introduce in figure 4.2 a grating made of M materials (numbered from 0 to M, separated by M-1 profiles (numbered from 1 to M). We introduce the following functions:

$$U_j = \begin{cases} F_j e^{-i\alpha_0 x} - F^i e^{-i\alpha_0 x} \delta_{j,M} & \text{in } V_j \\ 0 & \text{elsewhere.} \end{cases} \quad (4.43)$$

The function $U_j$ inside each layer can be expressed from the fields and normal derivatives on the lower and upper interface, for j=1, M-1:

$$\begin{aligned}
U_j(x,y) &= \int_{S_j} \mathcal{G}_j^+(x,y,s_j')\phi_j^+(s_j')ds_j' + \int_{S_j} \mathcal{N}_j^+(x,y,s_j')\psi_j^+(s_j')ds_j' \\
&- \int_{S_{j+1}} \mathcal{G}_{j+1}^-(x,y,s_{j+1}')\phi_{j+1}^-(s_{j+1}')ds_{j+1}' - \int_{S_{j+1}} \mathcal{N}_{j+1}^-(x,y,s_{j+1}')\psi_{j+1}^-(s_{j+1}')ds_{j+1}'.
\end{aligned} \quad (4.44)$$

The expression being limited to the second one if j = 0 and to the first one for j = M.

The functions derived from the Green functions in the various materials depend on the interface number:

$$\mathcal{G}_j^\pm(x,y,s') = \frac{1}{2id} \sum_{m=-\infty}^{\infty} \frac{1}{\gamma_{j,m}^\pm} \exp\left\{imK\left[x - x'(s_j')\right] + i\gamma_{j,m}^\pm \left|z - z'(s_j')\right|\right\}, \quad (4.45)$$

$$\mathcal{N}_j^\pm(x,y,s') = \frac{1}{2d} \sum_{m=-\infty}^{\infty} \left\{\frac{dx'(s')}{ds'}\text{sgn}\left[z - z'(s_j')\right] - \frac{\alpha_m}{\gamma_{j,m}^\pm}\frac{dz'(s_j')}{ds'}\right\} \times \\ \times \exp\left\{imK\left[x - x'(s_j')\right] + i\gamma_{j,m}^\pm \left|z - z'(s_j')\right|\right\}, \quad (4.46)$$

with:

$$\gamma_{j,m}^+ = \gamma_{j+1,m}^- = \sqrt{(kn_j)^2 - \alpha_m^2}. \quad (4.47)$$

It can be shown it the same manner as in eq. (4.146) that a compatibility condition on the j$^{th}$ interfacewritten in a matrix form is given by:

$$\frac{\psi_j^+}{2} = \mathcal{G}_j^+\phi_j^+ + \mathcal{N}_j^+\psi_j^+ - \mathcal{G}_{j,j+1}^-\phi_{j+1}^- - \mathcal{N}_{j,j+1}^-\psi_{j+1}^-. \quad (4.48)$$

Another compatibility equation is obtained on the (j+1)$^{th}$ interface:

$$\frac{\psi_{j+1}^-}{2} = \mathcal{G}_{j+1,j}^+\phi_j^+ + \mathcal{N}_{j+1,j}^-\psi_j^+ - \mathcal{G}_{j+1}^-\phi_{j+1}^- - \mathcal{N}_{j+1}^-\psi_{j+1}^-. \quad (4.49)$$

In eqs. (4.48) and (4.49), we use the double-index Greens functions that are derived on two consecutive profiles: $\mathcal{G}_{j,j+1}^- \equiv \mathcal{G}_j^-(s_j, s_{j+1}')$, $\mathcal{G}_{j+1,j}^+ \equiv \mathcal{G}_j^+(s_{j+1}, s_j')$, and similarly for $\mathcal{N}$.

The computability equation becomes, for the upper and lower media:

$$\frac{\psi_M^+}{2} = \mathcal{G}_M^+\phi_M^+ + \mathcal{N}_M^+\psi_M^+, \quad (4.50)$$

$$\frac{\psi_1^-}{2} = -\mathcal{G}_1^-\phi_1^- - \mathcal{N}_1^-\psi_1^-. \quad (4.51)$$

By combining eqs (4.48) with eq. (4.49), we obtain the link between the unknowns on the upper and on the lower interface of the j$^{th}$ layer, for j=1, M-1:

$$\begin{pmatrix} \mathcal{N}^-_{j,j+1} & \mathcal{G}^-_{j,j+1} \\ \mathcal{N}^-_{j+1} + \dfrac{\mathbb{I}}{2} & \mathcal{G}^-_{j+1} \end{pmatrix} \begin{pmatrix} \psi^-_{j+1} \\ \phi^-_{j+1} \end{pmatrix} = \begin{pmatrix} \mathcal{N}^+_j - \dfrac{\mathbb{I}}{2} & \mathcal{G}^+_j \\ \mathcal{N}^+_{j+1,j} & \mathcal{G}^+_{j+1,j} \end{pmatrix} \begin{pmatrix} \psi^+_j \\ \phi^+_j \end{pmatrix}. \quad (4.52)$$

This equation gives the transmission operator of the unknown amplitudes across the j$^{th}$ layer, i.e. from the j$^{th}$ to the (j+1)$^{st}$ interface, for j=1,M-1:

$$T_{j+1,j} = \begin{pmatrix} \mathcal{N}^-_{j,j+1} & \mathcal{G}^-_{j,j+1} \\ \mathcal{N}^-_{j+1} + \dfrac{\mathbb{I}}{2} & \mathcal{G}^-_{j+1} \end{pmatrix}^{-1} \begin{pmatrix} \mathcal{N}^+_j - \dfrac{\mathbb{I}}{2} & \mathcal{G}^+_j \\ \mathcal{N}^+_{j+1,j} & \mathcal{G}^+_{j+1,j} \end{pmatrix}. \quad (4.53)$$

The transmission operator includes an inverse operator. Numerically, this inversion leads to the inversion of a matrix, as we will see in section 4.6.

The transmission matrix across the j$^{th}$ interface for j = 1, …, M−1 is obtained using the continuity of the tangential and normal field components, as given by eqs.(4.8) and (4.10):

$$\begin{pmatrix} \psi^+_j \\ \phi^+_j \end{pmatrix} = T_j^{+-} \begin{pmatrix} \psi^-_j \\ \phi^-_j \end{pmatrix}, \qquad T_j^{+-} = \begin{pmatrix} \mathbb{I} & 0 \\ 0 & \dfrac{q^-_j}{q^+_j}\mathbb{I} \end{pmatrix}, \; j \neq M. \quad (4.54)$$

The advantage of this presentation is that there are no exponentially growing terms in the transmission matrices, since all the components of the two variable functions contain only scattered propagating or decreasing evanescent waves, in contrast with the other methods (differential, Fourier modal, Rayleigh, etc.). However, this formulation requires calculating the cross-layer functions between the interfaces, which leads to computation times equal to those of single-interface functions. As a consequence, the total computation time is almost multiplied by a factor 2 with respect to the case where cross-layer kernels can be avoided, as discussed in the next section.

Finally, it can be deduced from eqs. (4.53) and (4.54):

$$\begin{pmatrix} \psi^+_M \\ \phi^+_M \end{pmatrix} + \begin{pmatrix} \psi^i \\ \phi^i \end{pmatrix} = T^{+-}_M T_{M,M-1} \; \cdots \; T^{+-}_2 T_{2,1} T^{+-}_1 \begin{pmatrix} \psi^-_1 \\ \phi^-_1 \end{pmatrix}. \quad (4.55)$$

Finally, eqs. (4.55), (4.50) and (4.51) form an operator system of 4 equations with 4 unknown functions, which can be solved on a computer after representing each operator by a matrix, as described in section 6.

### 4.4.2. Profiles without interpenetration

This case is simpler than the situation in sec.4.1, because it is possible to use the plane-wave expansion between the grating profiles and thus to decouple the integral presentation used in eq.(4.44). The idea is illustrated in figure 4.3.

If it is possible to introduce a plane layer between the profiles, the plane wave expansion is valid inside this layer. The advantage is that the plane waves (propagating and evanescent) that represent each diffraction order m can be easily separated into to sets: (i) upgoing waves having amplitudes of the y-component of the field equal to $r_{j,m}$ that are generated by the lower surface $S_j$, (ii) downgoing waves with amplitudes $t_{j,m}$ generated by the upper grating surface $S_{j+1}$.

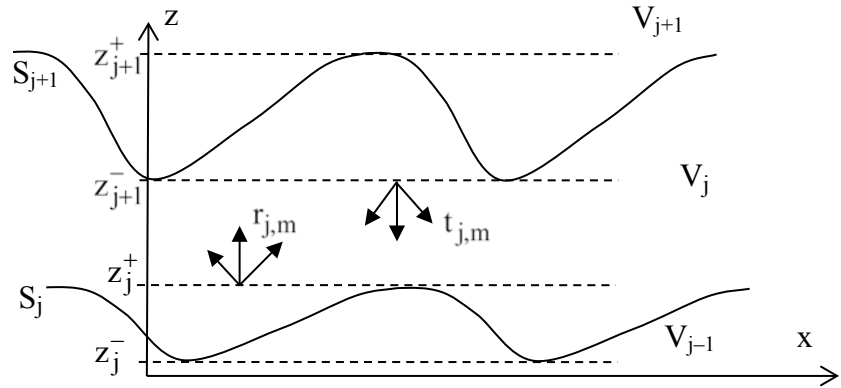

*Figure.4.3. Layer with two profiles that are separable by a plane layer*

Let us first rewrite eqs.(4.184) at $z > z_j^+$ :

$$r_{j,m} = \int_{s=0}^{S_{j,d}} \left[ N_{j,m}^+(s_j)\psi_j^+(s_j) + G_{j,m}^+(s_j)\phi_j^+(s_j) \right] ds_j , \qquad (4.56)$$

with

$$G_{j,m}^+(s_j) = \frac{1}{2id\gamma_{j,m}^+} \exp\left[ -imKx(s_j) - i\gamma_{j,m}^+ z(s_j) \right], \qquad (4.57)$$

$$N_{j,m}^+(s_j) = \frac{1}{2d}\left[ \frac{dx(s_j)}{ds_j} - \frac{\alpha_m}{\gamma_{j,m}^+}\frac{dz(s_j)}{ds_j} \right] \exp\left[ -imKx(s_j) - i\gamma_{j,m}^+ z(s_j) \right].$$

and with $s_j$ denoting the curvilinear abscissa on the $j^{th}$ profile, $s_{j,d}$ being the curvilinear abscissa of the point of $S_j$ of abscissa $x = d$.

We then can directly use eq.(4.48) to express the field on the interface $z = z(s_{j+1})$ :

$$\frac{\psi_{j+1}^-(s_{j+1})}{2} = \sum_m \exp\left[ imKx(s_{j+1}) + i\gamma_{j,m}^+ z(s_{j+1}) \right] r_{j,m}^+ - \mathcal{G}_{j+1}^- \phi_{j+1}^- - \mathcal{N}_{j+1}^- \psi_{j+1}^- . \qquad (4.58)$$

Let us consider the sum in eq. (4.58):

$$\zeta_{j+1} = \sum_m \exp\left[ imKx(s_{j+1}) + i\gamma_{j,m}^+ z(s_{j+1}) \right] r_{j,m} . \qquad (4.59)$$

Inserting in this equation the value of $r_{j,m}$ given by eq. (4.56) yields:

$$\zeta_{j+1}(s_{j+1}) = $$

$$\sum_m \exp\left[ imKx(s_{j+1}) + i\gamma^+_{j,m} z(s_{j+1}) \right] \int_{s_j=0}^{S_{j,d}} \left[ N^+_{j,m}(s_j) \psi^+_j(s_j) + G^+_{j,m}(s_j) \phi^+_j(s_j) \right] ds_j, \quad (4.60)$$

which can be written in operator form:

$$\zeta_{j+1} = \mathbf{N}^+_{j,j+1} \psi^+_j + \mathbf{G}^+_{j,j+1} \phi^+_j, \quad (4.61)$$

where the operators $\mathbf{N}^+_{j,j+1}$ and $\mathbf{G}^+_{j,j+1}$ are obtained from a summation in m and an integral in $s_j$. Thus we can write eq. (4.58) in the operator form:

$$\frac{\psi^-_{j+1}}{2} = \mathbf{N}^+_{j,j+1} \psi^+_j + \mathbf{G}^+_{j,j+1} \phi^+_j - \mathcal{G}^-_{j+1} \phi^-_{j+1} - \mathcal{N}^-_{j+1} \psi^-_{j+1}. \quad (4.62)$$

The second relation comes from eq. (4.49) by using the amplitudes of the diffraction orders coming down from the upper interfaces and valid below $z = z^-_{j+1}$. Following the same lines as in the previous paragraph yields:

$$\frac{\psi^-_j}{2} = \mathbf{N}^-_{j+1,j} \psi^-_{j+1} + \mathbf{G}^-_{j+1,j} \phi^-_{j+1} + \mathcal{G}^+_j \phi^+_j + \mathcal{N}^+_j \psi^+_j. \quad (4.63)$$

The transmission matrix between the $j^{th}$ and the $j+1^{st}$ interface takes a form similar to the case of interpenetrating layers, eq.(4.53). However, the difference is essential, because each series used in (4.63) is evaluated on a single interface:

$$T_{j+1,j} = \begin{pmatrix} \mathbf{N}^-_{j+1,j} & \mathbf{G}^-_{j+1,j} \\ \mathcal{N}^-_{j+1} + \frac{\mathbb{I}}{2} & \mathcal{G}^-_{j+1} \end{pmatrix}^{-1} \begin{pmatrix} \mathcal{N}^+_j - \frac{\mathbb{I}}{2} & \mathcal{G}^+_j \\ \mathbf{N}^+_{j,j+1} & \mathbf{G}^+_{j,j+1} \end{pmatrix}. \quad (4.64)$$

The second difference is that the exponential terms are explicitly given in the $\mathbf{N}^+_{j,j+1}$, $\mathbf{N}^-_{j+1,j}$, $\mathbf{G}^+_{j,j+1}$ and $\mathbf{G}^-_{j+1,j}$ through the functions $\exp\left[ imKx(s_{j+1}) + i\gamma^+_{j,m} z(s_{j+1}) \right]$ and $\exp\left[ imKx(s_j) - i\gamma^+_{j,m} z(s_j) \right]$ in such a way that it can be extracted from each of these operators a part containing all the growing and decreasing exponential terms, which allow a much better stability of the numerical implementation through adequate treatments, for example the S-matrix algorithm described in appendix A and B of Chapter 7.

### 4.5. Gratings in conical mounting

When classical gratings with one-dimensional periodicity are used with incidence plane perpendicular to the grooves, the diffraction orders lie in the same plane. Off-plane incidence brings the diffraction orders out of the plane and their directions lie on a cone, which explains the term of conical diffraction. One of the first experimental works can be found in [24, 25]. The use of conical mount is typical for concave gratings and in some spectrographs aiming to separate off-plane the incident and the diffracted directions.

The first theoretical studies were made in 1971 using the integral [26] and the differential [30] methods. An interesting theoretical development came in 1972 when Maystre and Petit demonstrated that under special conditions, mechanically ruled perfectly conducting gratings can have very high and constant efficiency over a large spectral domain [31]. They also established *the theorem of invariance* [32] that gives an expression of the diffraction efficiency of a perfectly conducting gratings in conical mounting, expressed as a linear combination of efficiencies in an in-plane (classical) mount for the two fundamental polarizations. Since the theorem is not valid for finitely conducting metals, later development of the integral method allowed studies of diffraction gratings behaviour in conical mount when working in the UV and visible [33, 34]. An interested reader can find the demonstration of the invariance theorem in [7, 33].

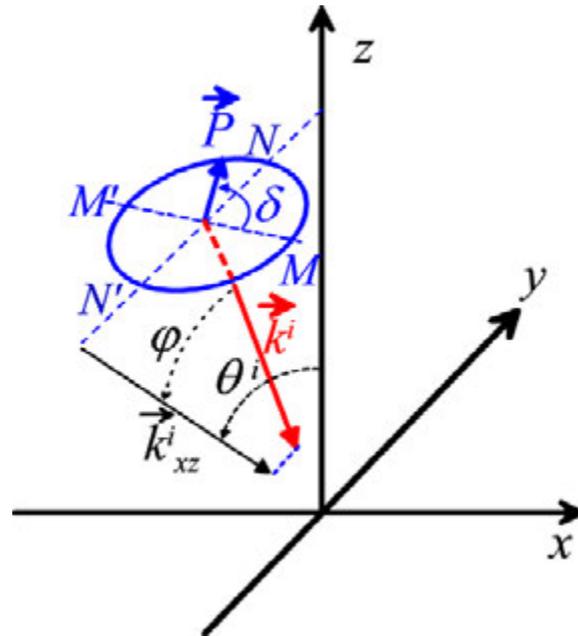

*Figure 4.4. Parameters of the incident wave in conical mount. The angle $\varphi$ denotes the angle between the incident wavevector $\vec{k}^i$ and its projection $\vec{k}^i_{xz}$ on the xz plane. The angle $\theta^i$ is the angle between the z axis and $\vec{k}^i_{xz}$. In order to define the polarization of the incident field, we construct the circle MNM'N' in the plane perpendicular to the incident wavevector $\vec{k}^i$, with the continuation of NN' intersecting the z axis and MM' being perpendicular to NN'. The polarization angle $\delta$ is the angle between M'M and the direction of the incident wavevector $\vec{k}^i$.*

The notations are summarized in figure 4.4. The mathematical formulation of the invariance theorem takes the form of an equivalence between the conical case and an associated classical case:
(i) conical case:

With incident angles $\theta^i$ and $\varphi$, incident polarization angle $\delta$, incident wavelength $\lambda$, the efficiencies in the various orders are denoted by $\rho_m(\theta^i, \varphi, \delta, \lambda)$.

(ii) fictitious equivalent classical case:

The wavelength is increased to $\lambda' = \lambda/\cos\varphi$, the angle $\varphi'$ is now equal to 0 (in-plane incidence), the angle $\theta'^i = \theta^i$, then efficiencies in TE and TM polarizations are denoted by $\rho_m^{TE}(\theta^i, \lambda')$ and $\rho_m^{TM}(\theta^i, \lambda')$, respectively. It can be noticed that the projection $\vec{k}^i_{xz}$ of the wavevector of the incident wave on the xz plane in conical mount identifies with the wavevector of the incident waves in the fictitious equivalent case.

The invariance theorem states that:

$$\rho_m\left(\theta^i,\varphi,\delta,\lambda\right) = \left(\cos\delta\right)^2 \rho_m^{TE}\left(\theta^i,\lambda'\right) + \left(\sin\delta\right)^2 \rho_m^{TM}\left(\theta^i,\lambda'\right). \tag{4.65}$$

It is to be noticed that, as for the incident wavevector, the projections of the wavevectors of the scattered waves on the xz plane in conical mount are identical to the wavevectors of the scattered waves in the fictitious equivalent case.

### 4.6. Numerical tools for an efficient numerical implementation

#### 4.6.1. Integration schemes for the integral equation

All the integral equations in this chapter link the value of an unknown function u(s) at a given point of S to its value on its entire domain of definition:

$$c\, u(s) = v(s) + \int_0^{s_d} W(s,s')u(s')ds', \tag{4.66}$$

where all functions are periodical, with v and W being known functions, W being possibly singular but integrable. The constant c takes values 0 or 1 according to whether the integral equation is of the first or second kind.

There are many different ways to solve such an equation for the grating problem. Pavageau et al. proposed an iterative method [13] that does not require any matrix inversion, like the well known Born method for scattering problems. Unfortunately, it may diverge [35, 36].

A well known general methods is based on the periodicity of all functions of the equation, which allows a projection of these functions and of the equation on the Fourier space:

$$u(s) = \sum_m u_m \exp(imK_s s), \quad K_s = \frac{2\pi}{s_d}, \tag{4.67}$$

and similar expressions for v and W. The integral equation is transformed into a linear system of algebraic equations:

$$\sum_m (W_{nm} - c\delta_{nm})u_m = v_n, \quad \forall n, \tag{4.68}$$

which can be solved numerically after truncation. However, this approach requires computing a double Fourier decomposition:

$$W_{nm} = \int_0^{s_d}\int_0^{s_d} W(s,s')u(s')\exp(-inK_s s - imK_s s')ds\, ds'. \tag{4.69}$$

The method has been applied to gratings with profiles consisting of few straight segments, because in that case the double Fourier integral can be calculated in closed form. It is so for triangular profiles [11] or trapezoidal gratings [37].

The most widespread method is the so-called point-matching (or discretization) method. Instead ot using discrete Fourier components, the unknown function is discretized on the grating profile and represented by its vales $u_j = u(s_j)$ inside the interval of integration. Ther integration process leads toan equation quite similar to eq.(4.68)

$$\sum_p \left( W_{jp} - c\delta_{jp} \right) u_p = v_j, \quad \forall j \in [1, P], \tag{4.70}$$

with P being the number of matching points. It is worth noting that the value of $W_{jp}$ may differ from the value of $W(s_j, s_p)$ obtained through the rectangular rule of integration (multiplied by the weight of integration) and can require much more complicated treatments, specially if W is singular. The sophistication of this treatment is one of the decisive keys for the precision of the solution of the integral solution. The second key is the analytical study of the kernel, which is described in the next two sections.

When W is regular, continuous, with a continuous first derivative, the rectangular rule of integration is quite precise since the function to be integrated is periodic and it can be noticed that the rectangular and trapezoidal rules are completely equivalent in that case. However, the derivative of W(s,s') is generally discontinuous when s = s' and a trapezoidal rule or higher order treatment provide a better precision [38, 39]. In what follows we assume that u(s) is a continuous function. This is obviously the case when the grating profile has no edges. Several more detailed arguments in favour of the rectangular rule can be found in [6, 7]. Let us shortly repeat one of them. We consider an integral of a periodical continuous function a(s):

$$a_0 = \int_0^{s_d} a(s) ds. \tag{4.71}$$

The exact integral is equal to the $0^{th}$ term in the Fourier expansion of a(s):

$$a(s) = \sum_m a_m \exp(imK_s s). \tag{4.72}$$

Simple calculations show that when using the rectangular rule with P discretization points, the integration error is proportional to the $P^{th}$ Fourier coefficient of a(x). Since it is continuous, the Fourier series converges like $1/P^2$ at least.

The implementation of the rectangular rule is very simple. We define the values of $W_{jp}$ in the following manner:

$$W_{j,p} = \frac{s_d}{P} W\left( \frac{j}{P} s_d, \frac{p}{P} s_d \right), \quad j, p = 1, \ldots, P-1. \tag{4.73}$$

Using this result, the product of the unknown functions u(s) with the non-singular parts of the kernels can be integrated in the form of a simple matrix product:

$$\int_0^{s_d} W(s_j, xs) u(s') ds' \approx \sum_p W_{j,p} u_p. \tag{4.74}$$

In the case of multilayer gratings with profiles interpenetration, there are two main situations that can complicate the numerical evaluation of the cross-layer functions which link the fields and normal derivatives on both sides of a layer:
(i)    The wavelength λ is much larger than the layer thickness t. In that case the profiles are located too close to each other, compared to λ, so that the functions $\mathcal{G}$ and $\mathcal{N}$ become large in modulus for $s_{j+1} \to s_j$. This behaviour has the same origin as the singularities of the kernels $\mathcal{G}$ and $\mathcal{N}$ for a bare grating, which will be discussed in section 6.3 and can be eliminated in a similar manner. Numerical results show no problems as long as layer thicknesses exceed $\lambda/20$.

(ii)     The period d is much greater than the layer thickness t. If the wavelength is not too much larger than t, then the kernels have no singularities, but their moduli present peaks when the distance $\left|P_{j+1,p}P_{j,q}\right|$ between two points located on the two different profiles is small with respect to the discretization distance between the points located on the same profile. The width of these maxima is of the order of magnitude of the layer thickness, thus the correct implementation the trapezoidal integration rule requires the distance between two consecutive points of the profile discretization $\Delta = \left|s_{j,p} - s_{j,p-1}\right|$ to be less than the width of the maxima. As a rule of thumb, if $\Delta \approx d/N_p$, where $N_p$ is the number of integration points, then its lower limit is determined by the relation:

$$N_{p,min} \propto \frac{d}{t}. \tag{4.75}$$

Thus, for echelles, a values of d of about 10 μm and t of 20 nm requires the number of integration points to exceed 500. Note that $N_p$ directly determines the number of unknown values of $\phi_j$ and $\psi_j$ and thus the size of the matrices to be used. Practically, it is not worth nowadays for $N_p$ to exceed 10 000, because of the computation time, memory requirements, round-off errors and limited digits. As a consequence, it is necessary to find another way of integration instead of the trapezoidal rule.

   There is another problem that can come from the matrix inversion in the construction of the transmission matrix between the layers, eq.(4.53). Contrary to the transmission matrix that contains growing and decreasing exponential terms in the plane wave expansion, (thus requires some type of recursive algorithm to contain the contamination of the growing exponentials, S-matrix algorithm, for example, see appendix 4.C), the distance between the profiles in the z-direction that appears in the kernels in eqs. (4.45) and (4.46) restricts the terms to only propagation or evanescently decreasing ones. However, the matrix inversion of these terms that is required in eq.(4.53) can create exponentially growing terms. Two situations can appear:
1. The matrix inversion in eq.(4.53) can be done without numerical problems. This happens when the layer thickness is not quite large. In that case it is possible to progress upwards in the stack of layers by following the S-matrix algorithm.
2. The matrix inversion does not work. This could happen if the distance between two consecutive interface is large. Two different geometries can be concerned:
(i)     there is no interpenetration of these two profiles. In that case one can easily apply the technique described in the next section.
(ii)     there is interpenetration of two very deep interfaces. It is possible to use directly eq.(4.52) in the S-matrix algorithm without inverting the matrix to calculate the entire T-matrix in eq. (4.53). The formulation of the S-matrix algorithm to an equation having the form given in (4.52) is quite similar to the classical aplication, but it requires one additional iteration step. The advantage is that it avoids the inversion of small terms that can lead to singular matrices. This special technique is given in Appendix 4.C and is not quite popular, but can be used in other methods that apply for multilayer stack, for example, in the coordinate transformation method.

### *4.6.2. Summation of the kernels*

There are several problems in the calculation of the functions included in eqs (4.19) and (4.23):

$$\mathcal{G}^{\pm}(s,s') = \frac{1}{2id} \sum_{m=-\infty}^{\infty} \frac{1}{\gamma_m^{\pm}} e^{imK[x(s)-x'(s')]+i\gamma_m^{\pm}|z(s)-z'(s')|}, \qquad (4.76)$$

$$\mathcal{N}^{\pm}(s,s') = \frac{1}{2d} \sum_{m=-\infty}^{\infty} \left[ \frac{dx'}{ds'} \operatorname{sgn}(z(s)-z'(s')) - \frac{\alpha_m}{\gamma_m^{\pm}} \frac{dz'}{ds'} \right] e^{imK[x(s)-x'(s')]+i\gamma_m^{\pm}|z(s)-z'(s')|}, \qquad (4.77)$$

$$\mathcal{K}(s,s') = \frac{1}{2d} \sum_{m=-\infty,+\infty} \left[ \operatorname{sgn}(z-z') \frac{dx}{ds} - \frac{\alpha_m}{\gamma_m^+} \frac{dz}{ds} \right] e^{imK[x(s)-x'(s')]+i\gamma_m^{\pm}|z(s)-z'(s')|}, \qquad (4.78)$$

with:

$$\alpha_m = \alpha_0 + mK, \qquad (4.79)$$

$$\gamma_m^{\pm} = \sqrt{(kn^{\pm})^2 - \alpha_m^2}. \qquad (4.80)$$

For the sake of simplicity, we assume here that $n^{\pm} = 1$ and we cancel the superscript $\pm$ in $\gamma_m^{\pm}$, $\mathcal{G}^{\pm}$ and $\mathcal{N}^{\pm}$.

Let us at first evaluate the asymptotic values of $\gamma_m$ and $\alpha_m$ for large values of m:

$$\begin{aligned} \alpha_m &\xrightarrow[m\to\infty]{} mK, \\ \gamma_m &\approx i|\alpha_m| - \frac{ik^2}{2|\alpha_m|} \xrightarrow[m\to\infty]{} i|\alpha_0 + mK|, \\ \frac{1}{\gamma_m} &\approx \frac{1}{i|\alpha_m|} + \frac{k^2}{2i|\alpha_m|^3} \xrightarrow[m\to\infty]{} \frac{1}{i|m|K}. \end{aligned} \qquad (4.81)$$

We consider the function

$$\mathcal{G}(s,s') = \frac{1}{2id} \sum_{m=-\infty}^{\infty} \frac{1}{\gamma_m} \exp\{imK[x(s)-x(s')] + i\gamma_m|z(s)-z(s')|\}. \qquad (4.82)$$

At point $s = s'$, it is obvious that the sum does not converge since the terms decrease in $1/|m|$. Of course, a very slow convergence can be expected when the two points are close to each other. Different techniques have been proposed to accelerate the convergence. Neureuther and Zaki [15] have employed a transformation technique based on the use of Mellin transforms. Other authors have proposed accelerating processes [40-43]. Here we describe a direct approach [7]. Let us determine at first the asymptotic expression of the kernel. If we replace (4.81) into eq.(4.82), we obtain the asymptotic term in the sum:

$$\mathcal{G}_\infty(s,s') = e^{\alpha_0[x(s)-x(s')]} \sum_{m=-1}^{-\infty} \frac{1}{4\pi m} e^{mK[x(s)-x(s')+i|z(s)-z(s')|]}$$
$$- e^{-\alpha_0[x(s)-x(s')]} \sum_{m=1}^{\infty} \frac{1}{4\pi m} e^{-mK[x(s)-x(s')-i|z(s)-z(s')|]}, \quad (4.83)$$

which contains two sums of the form $\sum_{m=1}^{\infty} \xi^m/m$ and can be summed in closed form:

$$\mathcal{G}_\infty(s,s') = \frac{1}{4\pi} e^{\alpha_0[x(s)-x(s')]} \log\left(1 - e^{-K[x(s)-x(s')+i|z(s)-z(s')|]}\right)$$
$$+ \frac{1}{4\pi} e^{-\alpha_0[x(s)-x(s')]} \log\left(1 - e^{-K[x(s)-x(s')-i|z(s)-z(s')|]}\right). \quad (4.84)$$

The calculation of the kernel in (4.82) is achieved by subtracting the asymptotic value:

$$\mathcal{G} = \mathcal{G}_\infty + \left(\mathcal{G} - \mathcal{G}_\infty\right). \quad (4.85)$$

The first term in the right-and side is explicitely given in (4.84). It is singular for s = s'e. This singularity is integrable and is be treated in the next section.

The term between parenthesis in eq.(4.85) is obtained by combining the terms in the sums in eqs.(4.82) and (4.83). As far as the second one is the asymptotic value of the former, the series converges, whatever the values of s and s'. Furthermore, it is possible to show that by combining the terms m and −m in the sum, we finally obtain a rapidly converging series, whose terms decrease as $m^{-3}$ when $s = s'$, as shown later in eq.(4.87) Moreover, in this case this series is continuous and its value is simply given by:

$$\left(\mathcal{G} - \mathcal{G}_\infty\right)\Big|_{s=s'} = \frac{1}{2id\gamma_0} + \sum_{m\neq 0}\left(\frac{1}{2id\gamma_m} + \frac{1}{4\pi|m|}\right). \quad (4.86)$$

Using the third identity of eq.(4.81), for large values of m, the term in the sum is equal to:

$$\frac{1}{2id\gamma_m} + \frac{1}{4\pi|m|} \rightarrow -\frac{k^2}{4d|\alpha_m|^3} \rightarrow -\frac{d^2}{8\pi\lambda^2}\frac{1}{|m|^3}. \quad (4.87)$$

Obviously, the singularity and the slow convergence of $\mathcal{G}$ have been carried out by introducing the series $\mathcal{G}_\infty$. Fortunaley, this singularity is logarithmic and thus integrable, as shown in the next section.

Let us now deal with the second function defined in eq. (4.77):

$$\mathcal{N}(s,s') = \frac{1}{2d}\sum_{m=-\infty}^{\infty}\left\{\frac{dx(s')}{ds'}\text{sgn}[z(s)-z(s')] - \frac{\alpha_m}{\gamma_m}\frac{dz(s')}{ds'}\right\}e^{imK[x(s)-x(s')]+i\gamma_m|z(s)-z(s')|}. \quad (4.88)$$

At the first glance, the term $\text{sgn}[z(s)-z(s')]$ suggests us that this function is not continuous for s' = s, at least if $\frac{dz'}{ds'} \neq 0$ for s' = s. To deal with this term, we proceed in the same way as in eq.(4.85), by introducing an asymptotic value $\mathcal{N}_\infty$ and we set now:

$$\mathcal{N} = \mathcal{N}_\infty + (\mathcal{N} - \mathcal{N}_\infty),  \qquad (4.89)$$

with

$$\mathcal{N}_\infty(s,s') = \frac{1}{2d} \operatorname{sgn}[z(s)-z(s')]$$

$$+ \frac{1}{2d}\left\{\frac{dx(s')}{ds'}\operatorname{sgn}[z(s)-z(s')] - i\frac{dz(s')}{ds'}\right\} e^{-\alpha_0[x(s)-x(s')]} \sum_{m=1}^{\infty} e^{-mK[x(s)-x(s')]+imK[z(s)-z(s')]} \quad (4.90)$$

$$+ \frac{1}{2d}\left\{\frac{dx(s')}{ds'}\operatorname{sgn}[z(s)-z(s')] + i\frac{dz(s')}{ds'}\right\} e^{\alpha_0[x(s)-x(s')]} \sum_{m=-1}^{-\infty} e^{mK[x(s)-x(s')]+imK[z(s)-z(s')]}.$$

The sums can be evaluated in a closed form:

$$\mathcal{N}_\infty(s,s') = \frac{1}{2d} \operatorname{sgn}[z(s)-z(s')]$$

$$+ \frac{1}{2d}\left\{\frac{dx(s')}{ds'}\operatorname{sgn}[z(s)-z(s')] - i\frac{dz(s')}{ds'}\right\} \frac{e^{-\alpha_0[x(s)-x(s')]}}{e^{K[x(s)-x(s')]-imK[z(s)-z(s')]}-1} \qquad (4.91)$$

$$+ \frac{1}{2d}\left\{\frac{dx(s')}{ds'}\operatorname{sgn}[z(s)-z(s')] + i\frac{dz(s')}{ds'}\right\} \frac{e^{\alpha_0[x(s)-x(s')]}}{e^{K[x(s)-x(s')]+imK[z(s)-z(s')]}-1}.$$

As noticed for $\mathcal{G}$, the term $\mathcal{N} - \mathcal{N}_\infty$ must be considered as a series each term of which is obtained by making the difference of the corresponding terms in the sums in eqs.(4.88) and (4.90). This series converges much more rapidly than the series in eq. (4.88) and it is continuous at s = s'.

The limit of $\mathcal{N}$ when $s' \to s$ can be determined calculating the limits of the two terms $\mathcal{N}_\infty$ and $(\mathcal{N} - \mathcal{N}_\infty)$. After tedious calculations, we can deduce that this limit exists and is given by:

$$\mathcal{N}(s,s) = \lim_{s' \to s} \mathcal{N}(s,s') = -\frac{dz}{ds}\left(\frac{i}{2\pi\alpha_0} + \frac{1}{2d}\sum_{m=-\infty}^{\infty}\frac{\alpha_m}{\gamma_m}\right) + \frac{1}{4\pi}\frac{\dfrac{d^2z}{ds^2}}{\left(\dfrac{dx}{ds}\right)^2 + \left(\dfrac{dz}{ds}\right)^2}. \qquad (4.92)$$

This interesting results established by Pavageau and Bousquet[44] is very important for the numerical applications, because it shows that $\mathcal{N}(s,s)$ contains a series that canverges like $1/m^3$ (after adding the terms with negative and positive values of m).

Let us notice that the second derivative of the profile function appears in eq.(4.92) and it clearly requires the continuity of the first derivative, i.e., the absence of edges.

The third kernel $\mathcal{K}(s,s')$, given by eq. (4.36):

$$\mathcal{K}(s,s') = \frac{1}{2d}\sum_{m=-\infty,+\infty}\left[\operatorname{sgn}[z(s)-z(s')]\frac{dx}{ds} - \frac{\alpha_m}{\gamma_m^+}\frac{dz}{ds}\right]e^{imK[x(s)-x(s')]+i\gamma_m^+|z(s)-z(s')|} \qquad (4.93)$$

is deduced from $\mathcal{N}(s,s')$ by replacing the derivatives $\dfrac{dx'}{ds'}$ and $\dfrac{dz'}{ds'}$ by $\dfrac{dx}{ds}$ and $\dfrac{dz}{ds}$ and its study is quite similar. After after tedious calculations it can be shown that:

$$\mathcal{K}(s,s) = \lim_{s' \to s} \mathcal{N}_0(s,s') = -\frac{dz}{ds}\left(\frac{i}{2\pi\alpha_0} + \frac{1}{2d}\sum_{m=-\infty}^{\infty}\frac{\alpha_m}{\gamma_m}\right) - \frac{1}{4\pi}\frac{\dfrac{d^2z}{ds^2}}{\left(\dfrac{dx}{ds}\right)^2 + \left(\dfrac{dz}{ds}\right)^2}. \tag{4.94}$$

### 4.6.3. *Integration of kernel singularities*

Clearly, the asymptotic part of $\mathcal{G}_\infty(s,s')$ in eq.(4.84) is singular when $s \to s'$. After some calculations, it can be found that

$$\lim_{s \to s'} \mathcal{G}_\infty(s,s') = \frac{1}{4\pi}\ln\left(K^2\left\{[x(s) - x(s')]^2 + [z(s) - z(s')]^2\right\}\right). \tag{4.95}$$

Noting that $x(s) - x(s') \approx (s-s')\dfrac{dx}{ds}$ and $z(s) - z(s') \approx (s-s')\dfrac{dz}{ds}$, eq.(4.95) yields:

$$\lim_{s \to s'} \mathcal{G}_\infty(s,s') = \frac{1}{2\pi}\left\{\ln(2\pi) + \frac{1}{2}\ln\left[\left(\frac{dx}{ds}\right)^2 + \left(\frac{dz}{ds}\right)^2\right] + \ln\frac{|s-s'|}{d}\right\}. \tag{4.96}$$

The first two terms represent regular parts that can be integrated by using the rectangular rule, as shown later. Unfortunately, $\ln\dfrac{|s-s'|}{d}$ is not periodic and the rectangular rule is very poor when applied to nonperiodic functions. It is possible to overcome this difficulty [7] by considering another function defined on $(0, d)$:

$$\tilde{\mathcal{G}}_\infty(s,s') = \frac{1}{2\pi}\left[\ln\frac{|s-s'|}{d} + \ln\left(1 - \frac{|s-s'|}{d}\right)\right], \quad s,s' \in (0,d), \tag{4.97}$$

which has the same singularity as $\dfrac{1}{2\pi}\ln\dfrac{|s-s'|}{d}$ in the interval $(0, d)$, because $d - |s-s'|$ never vanishes when $s,s' \in (0,d)$. The advantage of his new function is that it is continuous except on the singularity, and all its derivatives with respect to s are the same on $s' = 0$ and $s' = d$ and thus we can use the rectangular rule.

We perform the integration of $\mathcal{G}_\infty(s,s')$ by setting:

$$\int_0^{s_d}\mathcal{G}_\infty(s,s')\phi(s')ds' = \int_0^{s_d}\left[\mathcal{G}_\infty(s,s')\phi(s') - \tilde{\mathcal{G}}_\infty(s,s')\phi(s)\right]ds' + \int_0^{s_d}\tilde{\mathcal{G}}_\infty(s,s')\phi(s)ds'. \tag{4.98}$$

The term in the square bracket is non-singular and can be integrated using the rectangular rule, if we take into account that:

$$\lim_{s' \to s}\left[\mathcal{G}_\infty(s,s')\phi(s') - \tilde{\mathcal{G}}_\infty(s,s')\phi(s)\right] = \frac{1}{2\pi}\left\{\ln(2\pi) + \frac{1}{2}\ln\left[\left(\frac{dx}{ds}\right)^2 + \left(\frac{dz}{ds}\right)^2\right]\right\}\phi(s). \tag{4.99}$$

The second term in eq.(4.98) contains the singular part, but it can be integrated analytically:

$$\int_0^{s_d} \tilde{\mathcal{G}}_\infty(s,s')\phi(s)ds' = \phi(s)\int_0^{s_d} \tilde{\mathcal{G}}_\infty(s,s')ds' = -\frac{d}{\pi}\phi(s). \qquad (4.100)$$

In conclusion, the integration of the singular kernel is made by introduction at first $\mathcal{G}_\infty$, which permits to define a series $\mathcal{G} - \mathcal{G}_\infty$ that is continuous and rapidly converging hence easily integrable by the use of the rectangular rule. Second, the integration of the term containing $\mathcal{G}_\infty$ is performed by defining a new function $\tilde{\mathcal{G}}_\infty$, which has the same singularity as $\mathcal{G}_\infty$, has the property of a periodic function, and can be analytically integrated.

### *4.6.4. Kernel singularities for highly conducting metals*

When the conductivity of a metallic grating tends to infinity, the Green function in the metal tends to a delta function. This property is rather obvious: for very large conductivities, the field generated by a line current (delta function) placed in the metal decreases very rapidly since it is absorbed on very short distances. This behaviour have drastic consequences on the kernels of the integral equations dealing with metallic gratings, which are directly derived from the Green function: the two variable functions relative to the metallic part of the grating tend to delta functions as well. The integration of such functions through a point matching method requires more and more points of discretization around $s' = s$ and, since s can take any value in the interval $(0, s_d)$ the integration and the inversion of the final linear system of equations (bearing in mind that its size is the total number of discretization points) becomes impossible. This remarks explains why the first attempts at implementing the integral equations on computers were not able to give any result for metallic gratings in the visible and infrared regions.

A very efficient way to overcome this difficulty is to apply an approach called *local summation* [7], using another form of the Green function [45]:

$$\mathcal{G}(\vec{r} - \vec{r}') = \frac{1}{4i}\sum_m e^{imd\alpha_0} H_0^+\left(k|\vec{r} - \vec{r}'| - md\,\hat{x}\right), \quad \vec{r} = (x, z), \qquad (4.101)$$

with $\hat{x}$ being the unit vector of the x axis. This form is the direct consequence of the fact that $\frac{1}{4i}H_0^+\left(k|\vec{r} - \vec{r}'|\right)$ is the Green function of the Helmholtz equation:

$$\nabla^2\left[\frac{1}{4i}H_0^+\left(k|\vec{r} - \vec{r}'|\right)\right] + k^2\left[\frac{1}{4i}H_0^+\left(k|\vec{r} - \vec{r}'|\right)\right] = \delta(\vec{r} - \vec{r}'). \qquad (4.102)$$

Since the pseudo-periodic Green function $\mathcal{G}(\vec{r} - \vec{r}')$ in vacuum is defined by:

$$\nabla^2\tilde{\mathcal{G}}(\vec{r},\vec{r}') + k^2\tilde{\mathcal{G}}(\vec{r},\vec{r}') = \sum_{m=-\infty,+\infty} e^{i\alpha_m d}\delta(\vec{r} - \vec{r}' - md\,\hat{x}), \qquad (4.103)$$

it follows that $\mathcal{G}(\vec{r} - \vec{r}')$ is a sum of Green functions $\frac{1}{4i}H_0^+\left(k|\vec{r} - \vec{r}'| - md\,\hat{x}\right)$ satisfying:

$$\nabla^2 \left[ \frac{1}{4i} e^{i\alpha_n d} H_0^+ \left(k|\vec{r}-\vec{r}\,'-md\hat{x}|\right) \right] + k^2 \left[ \frac{1}{4i} e^{i\alpha_n d} H_0^+ \left(k|\vec{r}-\vec{r}\,'-md\hat{x}|\right) \right] =$$
$$= e^{i\alpha_n d} \delta(\vec{r}-\vec{r}\,'-md\hat{x}). \tag{4.104}$$

Inserting the value of $\mathcal{G}(\vec{r}-\vec{r}\,')$ given by eq. (4.101) inside the integral $\int_0^{s_d} \mathcal{G}(s,s')\phi(s')ds'$, then making the change of variable $\vec{r}\,'-md\hat{x}=\vec{r}\,''$, and finally gathering the infinite sum of integrals on one period into a single integral from $-\infty$ to $+\infty$ yields:

$$\int_0^{s_d} \mathcal{G}(s,s')\phi(s')ds' = \frac{1}{4i} \int_{-\infty}^{\infty} H_0^+(k|\vec{r}-\vec{r}\,'|)e^{i\alpha_0(x'-x)}\phi(s')ds'. \tag{4.105}$$

In the same way it can be shown that

$$\int_0^{s_d} \mathcal{N}(s,s')\psi(s')ds' = \frac{ik}{4} \int_{-\infty}^{\infty} H_1^+(k|\vec{r}-\vec{r}\,'|)e^{i\alpha_0(x'-x)} \frac{z'-z-\frac{dz}{ds'}(x'-x)}{|\vec{r}-\vec{r}\,'|}\psi(s')ds'. \tag{4.106}$$

When the permittivity becomes very large in modulus, the functions $\mathcal{G}^-(s,s')$ and $\mathcal{N}^-(s,s')$ are obtained by replacing k by $kn^-$ in $\mathcal{G}(s,s')$ and $\mathcal{N}(s,s')$. Since the value of $n^-$ is close to an imaginary number in the visible and infrared regions for usual metals (for example, $n^- =$ 1.3 +i 7.11 for aluminum at 650 nm), the Hankel functions $H_0^+(k|\vec{r}-\vec{r}\,'|)$ and $H_1^+(k|\vec{r}-\vec{r}\,'|)$ become very close to the modified Bessel functions $K_0\left[k|n^-(\vec{r}-\vec{r}\,')|\right]$ and $K_1\left[k|n^-(\vec{r}-\vec{r}\,')|\right]$ (see [45]) and tend to delta functions when $|n^-|\to\infty$. Thus, these functions vary much more rapidly than the unknown function $\phi$, which can be considered as a constant. Thus, remarking that when $x' \approx x$

$$|\vec{r}-\vec{r}\,'| \approx |x-x'|\sqrt{1+\left(\frac{dz}{dx}\right)^2},$$
$$z'-z-(x'-x)\frac{dz}{dx} \approx -\frac{1}{2}\frac{d^2z}{dx^2}(x'-x)^2, \tag{4.107}$$
$$\exp[i\alpha_0(x'-x)] \approx 1,$$

yields finally:

$$\int_0^{s_d} \mathcal{G}(s,s')\phi(s')ds' \approx \frac{\phi(s)}{4i} \int_{-\infty}^{\infty} H_0^+(k|n^-(\vec{r}-\vec{r}\,')|)e^{i\alpha_0(x'-x)}ds' \approx \frac{\phi(s)}{2ik\sqrt{1+\left(\frac{dz}{dx}\right)^2}}, \tag{4.108}$$

and

$$\int_0^{s_d} \mathcal{N}(s,s')\psi(s')ds' \approx \frac{\frac{d^2z}{dx^2}\psi(s)}{4ik\left[1+\left(\frac{dz}{dx}\right)^2\right]^{3/2}} . \qquad (4.109)$$

It is not surprising to note that in this approximation, the calculations of the integrals require neither summation of the kernels, nor integration of the singularities, nor matrix multiplication. As the kernels tend toward delta-distributions with amplitudes determining the coefficients in eqs.(4.108) and (4.109), their matrix representations tend to diagonal matrices.

Numerical results have shown that this simpler formulation not only successfully applies in the domain where the summation and integration processes defined in the previous sections fail, but also remains valid with a good accuracy (about $10^{-3}$ in relative value) in a large domain of metals and wavelengths. For example, with aluminum, this approach works in the visible, a domain in which the classical method of integration can be used as well (but with a greater computation time), wheras the local summation is necessary for metals in the far-infrared and microwaves domain. It is very important to notice that, using the local summation and assuming that $|n^-|\to\infty$, it can be shown that the integral equation for metallic gratings described in appendix 4.B tends towards the integral equations obtained for perfectly conducting metals.

### *4.6.5. Problems of edges and non-analytical profiles*

When the grating profile has edges or corners (in 2D case), fundamental difficulties appear. First, the uniqueness of the solution of the electromagnetic field is not ensured. The hypothesis of the integrability of the unknown function u(x) in the integral equation is equivalent to the Meixner condition of integrability [46], although it is singular.

The second problem lies in the validity of the boundary condition of electromagnetic field on the grating profile. Indeed, on the edges, the normal and tangential direction on the profile are not defined in a unique manner. Moreover, when establishing these boundary conditions, the demonstration does not work if the surface has edges; However, these warnings are not dramatic. The demonstration of Archimede theorem is also questionable if the object presents edges. Does it exist any doubt about the validity of this theorem?

The third problem in the integral method lies in the process of integration in the vicinity of the edges. The kernels become discontinuous or even meaningless, depending on whether the point of calculation of the integral coincides with the edge point or not, see eq.(4.90) - (4.92). Moreover, the integration can fail due to the eventual singularity of the unknown function on the edge [44].

To overcome the edge problem, there exist several approaches. The most direct one consists in replacing the actual profile z = f(x) with its truncated Fourier representation:

$$\tilde{f}_M(x) = \sum_{m=-M}^{M} f_m e^{imKx} . \qquad (4.110)$$

The new profile has no edges and in most cases of ruled or holographic gratings, it mimics quite well the true profile. Numerical tests have shown a very good convergence of results with respect to the number 2M+1 of Fourier terms and the number P of discretization points, provided the empirical rule P > 4(2M + 1). This method can be used to describe some profiles that are not represented by continuous functions, for example lamellar gratings, provided that M is larger than 10. Unfortunately, there are several important cases of classical gratings and classes of relatively new types of gratings and periodic systems that cannot be treated using this simple approach.

The first problem covers the case of echelle gratings, working in grazing incidence in high and very high orders (see Chapter 1). The fact that the period is some tens or hundreds times larger than the wavelength $\lambda$, and that the working facet is quite steep (sometimes going up to 86° groove angle), requires that its geometry is represented in the method with an error smaller than $\lambda/20$. Simple but *incorrect* estimations show that this rule of thumb would be acceptable for numerical treatment: if the period is close to 50 wavelengths, we need about 1000 points of discretization and thus 250 Fourier harmonics, according to the rule $P > 4(2M + 1)$. However, Gibbs phenomenon will significantly modify the form of the working facet, which could be almost vertical. In order to correctly describe the field on this facet, we need to have at least 5 to 10 points per wavelength *along* the the working facet, rather than along its projection on the x-axis. With a 85° groove angle, the length of this facet is about 11 times its x-projection, in such a way that the number of discretization points must be multiplied by a factor 10. The number of Fourier components in the profile follows the same rule.

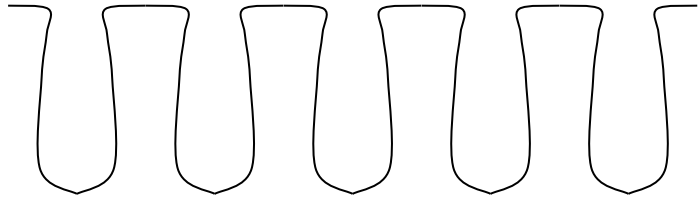

*Figure 4.5. Schematic representation of an etched grating profile with non-analytical function description.*

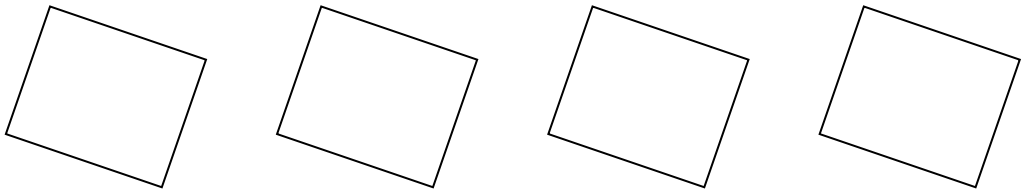

*Figure 4.6. Grating made of inclined rectangular cross-section rods.*

Another class of problems consists of unconventional geometries, like inverted slope grooves, obtained during groove etching technologies, as seen in Figure.4.5, or rod gratings, shown in Figure.4.6. We have noticed that the problem of vertical segments adds difficulties. For example, the problem of edges cannot be solved any longer by a Fourier expansion of the profile. It exists a possibility to simultaneously solve the two problems by introducing a curvilinear coordinate that follows the grating profile. As far as the integration is made along the profile, this curvilinear integration comes as a natural way of calculating the integrals in the integral equations. Moreover, adaptive meshing can be used to reduce the influence of edges.

In the real life, edges do not exist. On each edge there is at least one atom that has no edges, even though the light has a wavelength much larger than the atom dimensions. Edges of nuclear particles are not discussed even in the most exotic theories. Thus the idea is to replace the edges by arcs, where the partial derivatives can be well defined till the second order, which is sufficient for integral equations, as remarked in section 6.2. An adaptive density of discretization points gives the possibility to significantly increase the density of the points close to the initial edges, and not elsewhere. Let us, for example, consider a rod grating having a straight rectangular cross section. The segments 1 and 3 are parallel to Ox, the segments 2 and 4, to Oz. Let us assume the origin of the curvilinear coordinate at the bottom-left corner (figure 4.7).

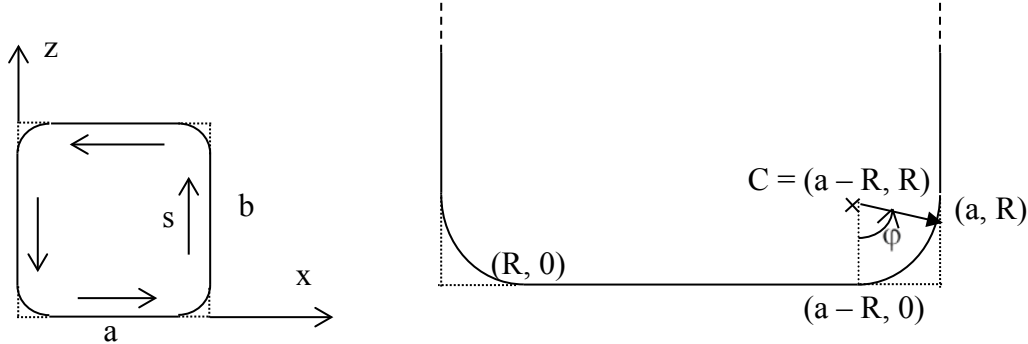

*Figure 4.7. Left: rounding of corners of a rectangular cross-section rod with side lengths a and b, together with x, y, and s coordinate lines. Right: schematic representation of different straight and arc segments to obtain the links between the Cartesian and the curvilinear coordinates.*

The coordinate s starts at $x = R$ and $z = 0$, and follows the interface along its different parts:

1) the first horizontal segment, $0 < s < a - 2R$:

$$x(s) = R + s,$$
$$z(s) = 0. \qquad (4.111)$$

Here the final value of s is equal to $s_{1,max} = a - 2R$:

2) the circular rounding of the bottom-right corner, defined by the equation:

$$[x - (a - R)]^2 + (z - R)^2 = R^2. \qquad (4.112)$$

Here, $0 < s - s_{1,max} < R\frac{\pi}{2}$:

$$s = R\varphi + s_{1,max},$$
$$x = a - R + R\sin(\varphi) = a - R + R\sin\frac{s - s_{1,max}}{R}, \qquad (4.113)$$
$$z = R - R\cos(\varphi) = R - R\cos\frac{s - s_{1,max}}{R}.$$

Here, $s_{2,max} = R\frac{\pi}{2} + s_{1,max}$

3) the next vertical segment at $x = a$, $0 < s - s_{2,max} < b - 2R$:

$$x = a,$$
$$z = s - s_{2,max} + R, \qquad (4.114)$$

and the same for the rest of the profile. The process is straightforward and needs an adapted application for each class of profiles. The advantage is that the derivatives $\frac{dx}{ds}$ and $\frac{dz}{ds}$ exist and are continuous everywhere on the profile. Thus, the second-order derivatives, which are required for the explicit summation and integration of the kernels, exist at least piesewisely, too. This could easily be checked at the point (a − R, 0), for example. For $s < s_{1,max}$, we use eqs.(4.111):

$$\frac{x(s)}{ds} = 1; \quad \frac{z(s)}{ds} = 0. \tag{4.115}$$

For $s_{1,max} < s < s_{2,max}$ it is necessary to use eqs.(4.113):

$$\frac{dx(s)}{ds} = \cos\frac{s - s_{1,max}}{R} = 1 \quad \text{for } s = s_{1,max},$$
$$\frac{dz(s)}{ds} = \sin\frac{s - s_{1,max}}{R} = 0 \quad \text{for } s = s_{1,max}. \tag{4.116}$$

The comparison of (4.115) and (4.116) points out the existence and continuity of the first derivatives.

It is worth noting that, in contrast with the adaptive *spatial* resolution used in several other methods (FEM, FDTD, RCW, coordinate transformation methods), here it is more convenient to call it adaptive *profile* resolution method, as far as it represents a 1D curvilinear coordinate adaptation.

Let us impose the requirement that the arc segments require $N_{arc}$ times larger density points than on the straight segments. The entire length along s of the profile in figure.4.7 (left) is equal to $L_{tot} = 2a + 2b − 8R + 2\pi R$. The total number of discretization points is related to the length ot the segments, R, and to $N_{arc}$. If the distance between the points along the straight segments is $\Delta$, it will be equal to $\Delta / N_{arc}$ on the arcs. Thus the total number of points for a single-rod per period is equal to:

$$P = \frac{2a + 2b - 8R + 2\pi R N_{arc}}{\Delta}. \tag{4.117}$$

In practice, unless some automatic scheme of determining the technical parameters of the computation is used, this equation determines $\Delta$ when the total number of integration points is chosen.

Starting from s = 0 in figure 4.8 (right), the abscissa values along s of the points on the interface are given by consecutively adding the values of the point number j to the end values of the previous segment:

1) first horizontal segment, $0 < s < s_{1,max}$ : $s_j = s_{j-1} + \Delta$,

2) circular rounding of the bottom-right corner, $s_{1,max} < s < s_{2,max}$ : $s_j = s_{j-1} + \frac{\Delta}{N_{arc}}$,

3) next vertical segment at x = a, $s_{2,max} < s < b - 2R + s_{2,max}$ : $s_j = s_{j-1} + \Delta$,

4) so on to close the rod, then eventually going to another object inside the same grating period.

An additional improvement can be performed by making a smooth transfer from $\Delta$ to $\Delta/N_{arc}$, i.e., to smoothly go from the density defined on the straight segments and on the arcs. This could be important for large-period systems with respect to the wavelength, if there is a restriction in the total number of points in the profile discretization. If so, we can lay on the fact that the smaller the curvature of the segment (i.e., the larger its length), the smoother the behaviour of the kernels and of the unknown functions. Thus in the middle of a straight segment we will place points that are more distant to each other than close to the extremeties of the segments. Maystre has adapted such approach in his code Grating 2000, by defining a specific distance along the large segments starting from their edges, so that the density of the discretization points increases when approaching the ends of the segments. This approach gave the possibility to model echelle gratings working in very high diffraction orders (600 or more) in the '90s, when no alternative approach was able to provide reliable results. The method was unbeatable for echelles, before Li and Chandezon [47] formulated an improvement of the coordinate transformation method to work for profiles with edges. However, the latter does not apply to rod gratings, or to profiles having the form as given in figure.4.5.

## 4.7. Examples of numerical results

All the results shown in this section are obtained using the code Grating 2000.

### *4.7.1. Sinusoidal perfectly conducting grating*

Table 1 shows the efficiencies in the two non-evanescent orders ($-1^{st}$ and $0^{th}$) of a sinusoidal perfectly conducting grating of period 600 nm and height 180 nm (from the bottom to the top of the groove) illuminated under incidence angle 30° and wavelength 600 nm. In this case, the $-1^{st}$ order is scattered with an angle of scattering (measured anticlockwise, in contrast with incident angle) of -30°, which entails that it propagates just in the direction which is the opposite to that of the incident wave (this is called Littrow mounting by specialists of gratings).

The number of discretisation points is P and the series included in the kernel are summed from $-M$ to $+M$. The symbol $\sum \rho_m$ denotes the sum of the two efficiencies, the energy balance being satisfied when $\sum \rho_m = 1$.

|        | TE polarization |          |                | TM polarization |          |                |
|--------|-----------------|----------|----------------|-----------------|----------|----------------|
| P,M    | $\rho_{-1}$     | $\rho_{-0}$ | $\sum \rho_m$ | $\rho_{-1}$     | $\rho_{-0}$ | $\sum \rho_m$ |
| 4,2    | 0.4658          | 0.5437   | 1.0095         | 0.9466          | 0.0514   | 0.9980         |
| 6,3    | 0.4703          | 0.5288   | 0.9991         | 0.9581          | 0.0411   | 0.9992         |
| 25,10  | 0.4669          | 0.5336   | 1.0005         | 0.9579          | 0.0421   | 1.0000         |
| 50,20  | 0.4659          | 0.5334   | 0.9993         | 0.9579          | 0.0421   | 1.0000         |
| 110,50 | 0.4659          | 0.5341   | 1.0000         | 0.9579          | 0.0421   | 1.0000         |

*Table 4.1. The sinusoidal perfectly conducting grating.*

It is worth noting that a precision better than 0.01 is reached as soon as P > 4 and M > 2!. A precision of $10^{-3}$ needs P > 50 and M > 20.

### *4.7.2. Echelette perfectly conducting grating*

The echelette grating is a grating with triangular groove (figure 4.8). The blaze angle b (angle of the large facet with the x axis) is equal to 30° and the apex angle A (angle between the two facets) to 90°. The other parameters are the same as in section 7.1. It must be noticed that the incidence angle and the blaze angle are equal, which entails that the incident wavevector is orthogonal to the large facet. In these conditions, it can be shown that for TM polarization, the grating problem can be solved in closed form: the efficiency in the -1$^{st}$ order is equal to unity while in the 0$^{th}$ order it vanishes [48].

The demonstration is straightforward: the sum of the incident wave and of a plane wave with unit amplitude propagating in the opposite direction satisfies all the conditions of the boundary value problem stated in section 3.2. The reader can notice that this sum satisfies the Helmholtz equation. In addition, this sum of two plane waves propagating in opposite directions constitute an interference system that presents white areas and dark lines. The maximum of white lines coincide with the large facets of the grating, and on these lines, the derivative of the field (thus the normal derivative) vanishes. The normal derivative with respect to the small facet vanishes as well since the field is invariant in the normal direction. At the first glance, this property is obvious since the field is "reflected" by the large facets, or in other words, the scattering phenomenon reduces in that case to a simple reflection phenomenon. This reasoning fails since the same phenomenon is not observed for TE polarization: it is dangerous to invoke reflection phenomena on the large facets when the width of these facets has the same order of magnitude as the wavelength of the light! The concentration of the incident energy in a single order is called 'blazing effect' and the theorem of Marechal and Stroke is the origin of the name 'blazed gratings' given sometimes to ruled gratings.

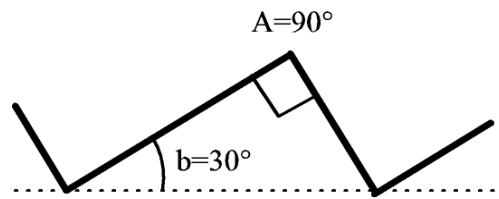

*Figure 4.8: A ruled grating*

| P,M | TE polarization | | | TM polarization | | |
|---|---|---|---|---|---|---|
| | $\rho_{-1}$ | $\rho_{-0}$ | $\sum \rho_m$ | $\rho_{-1}$ | $\rho_{-0}$ | $\sum \rho_m$ |
| 6,3 | 0.6531 | 0.4451 | 1.0982 | 1.4452 | 0.0012 | 1.4464 |
| 25,10 | 0.5838 | 0.4123 | 0.9961 | 0.9976 | 0.0001 | 0.9977 |
| 50,20 | 0.5838 | 0.4123 | 0.9961 | 0.9976 | 0.0001 | 0.9977 |
| 110,50 | 0.5929 | 0.4055 | 0.9984 | 0.9984 | 0.0000 | 0.9984 |
| 250,100 | 0.5932 | 0.4035 | 0.9967 | 0.9989 | 0.0000 | 0.9989 |

*Table 4.2. The echelette perfectly conducting grating and the Marechal and Stroke theorem.*

Table 4.2 shows that the convergence is significantly slower than for the sinusoidal grating, due to the edges. Nevertheless, a precision of about 0.01 is obtained when P > 25 and M > 10. With the same values, the Marechal and Stroke theorem is satisfied with a precision better than 0.003. The computation time is always less than 1second on a PC computer except for the last line, for which 2 seconds are required.

### 4.7.3. Lamellar perfectly conducting grating

The profile of a lamellar grating is shown in figure 4.9. The widths of the hole and of the bump are denoted by t and b and the height by h. In this example, b = t = 300 nm, and h = 180 nm.

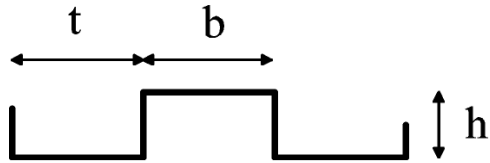

*Figure 4.9. A lamellar grating*

|  | TE polarization | | | TM polarization | | |
|---|---|---|---|---|---|---|
| P,M | $\rho_{-1}$ | $\rho_{-0}$ | $\sum \rho_m$ | $\rho_{-1}$ | $\rho_{-0}$ | $\sum \rho_m$ |
| 6,3 | 0.7770 | 0.6999 | 1.4769 | 23.66 | 5.09 | 28.75 |
| 25,10 | 0.3490 | 0.6394 | 0.9884 | 0.8293 | 0.1724 | 1.0016 |
| 50,20 | 0.3347 | 0.6569 | 0.9915 | 0.8191 | 0.1705 | 0.9896 |
| 150,70 | 0.3279 | 0.6659 | 0.9938 | 0.8201 | 0.1718 | 0.9919 |
| 250,100 | 0.3288 | 0.6733 | 1.0021 | 0.8214 | 0.1725 | 0.9939 |

*Table 4.3. The echelette perfectly conducting grating and the Marechal and Stroke theorem*

The main conclusion to draw from Table 4.3 is that the convergence for lamellar gratings is even slower than for echelette gratings. This is not surprising if we notice that the number of edges is multiplied by 2. A precision of about 0.02 is obtained when P > 50 and M > 20. The results for P = 6 and M = 3 are aberrant, specially for TM polarization.

### 4.7.4. Aluminum sinusoidal grating in the near infrared

We consider a sinusoidal aluminum grating with period d = 400 nm, a height h = 100 nm, illuminated with incidence angle 10° and wavelength 300 nm. With these parameters, three plane waves ($-1^{st}$, $0^{th}$ and $+1^{st}$) are reflected. The optical index of aluminum at 300 nm is equal to 4.2+ i 21.5. We give in Table 4.4 the efficiency in the $-1^{st}$ order and the sum of the three efficiencies.

|  | TE polarization | | TM polarization | |
|---|---|---|---|---|
| P,M | $\rho_{-1}$ | $\sum \rho_m$ | $\rho_{-1}$ | $\sum \rho_m$ |
| 6,3 | 0.5205 | 0.9582 | 0.4367 | 0.9618 |
| 10,4 | 0.5201 | 0.9649 | 0.4325 | 0.9521 |
| 30,13 | 0.5203 | 0.9655 | 0.4321 | 0.9518 |
| 100,45 | 0.5204 | 0.9655 | 0.4320 | 0.9518 |

*Table 4.4. The aluminum sinusoidal grating*

Table 4.4 shows a very good convergence of the results, similar to the convergence observed for sinusoidal perfectly conducting gratings, thanks to the local summation of the two variable functions of the kernel derived from the Green function in aluminum.

### *4.7.5. Buried echelette silver grating in the visible.*

The buried silver grating is shown in figure 4.10. A symmetric echelette silver grating with period 900 nm has been covered by a dielectric of index 1.5, the maximum depth e of dielectric being equal to 800 nm. This grating is illuminated in normal incidence .by a plane wave of wavelength 600 nm. The index of siver for this wavelength is equal to 0.006 + i 3.75. Table 4.5 gives the efficiency in the $0^{th}$ order and the total of efficiencies of the three reflected orders ($-1^{st}$, $0^{th}$ and $+1^{st}$).

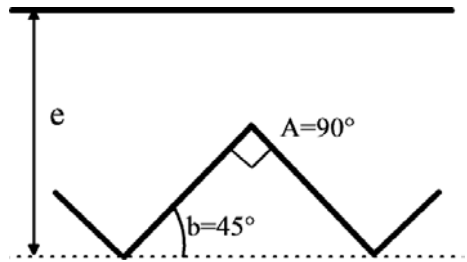

*Figure 4.10. A symmetric silver grating coated by dielectric*

|       | TE polarization |              | TM polarization |              |
|-------|-----------------|--------------|-----------------|--------------|
| P,M   | $\rho_0$        | $\sum \rho_m$ | $\rho_0$        | $\sum \rho_m$ |
| 10,4  | 0.4892D         | 0.9830       | 0.6316          | 1.0622       |
| 30,13 | 0.5164          | 0.9651       | 0.5665          | 0.9339       |
| 100,45| 0.5153          | 0.9600       | 0.6062          | 0.9178       |
| 200,90| 0.5153          | 0.9593       | 0.6168          | 0.9153       |

*Table 4.5. The buried silver grating*

The convergence is slower than in the preceding case and the results for P = 10 and M = 4 are not correct, specially for TM polarization. For TE polarization, a convergence of the results with a precision better than 0.006 is obtained for P = 30 and M = 13. This is not so for TM polarization, in which it is necessary to reach P = 100 and M = 45 to get a precision of the order of 0.003. Here, the method of local summation is not used, but this fact does not explain the slower convergence since the modulus of the optical index is not large. The main reason can be found in the edges of the profile. The slower convergence for TM polarization is rather general for metallic gratings, due to the existence of plasmon resonances on the grating surface.

### *4.7.6. Dielectric rod grating.*

The grating is made of dielectric elliptic rods with optical index 1.4, width w = 600 nm and height h = 400 nm (figure 4.11). The period is equal to 800 nm. It is illuminated with incidence 20° by a plane wave with wavelength 600 nm. Two orders ($-1^{st}$ and $0^{th}$) are reflected and transmitted. We give in Table 4.6 the efficiency in the $0^{th}$ transmitted order and the sum of efficiencies of the 4 scattered orders, which should be equal to 1 for a perfect energy balance.

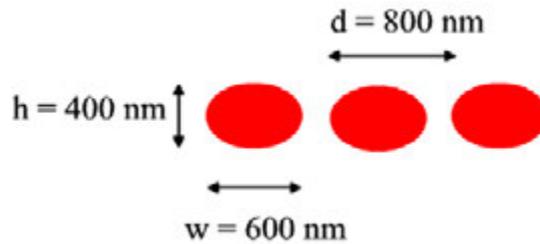

*Figure 4.11. A dielectric rod grating*

|  | TE polarization |  | TM polarization |  |
|---|---|---|---|---|
| P,M | $\tau_0$ | $\sum \rho_m + \sum \tau_m$ | $\tau_0$ | $\sum \rho_m + \sum \tau_m$ |
| 10,4 | 0.1824D | 0.9139 | 0.7893 | 1.2489 |
| 30,13 | 0.2163 | 0.9961 | 0.8161 | 1.0054 |
| 100,45 | 0.2073 | 1.0005 | 0.8187 | 1.0004 |
| 200,90 | 0.2070 | 1.0001 | 0.8187 | 1.0001 |

*Table4. 6. The dielectric rod grating with elliptic rods*

The precision for P = 30 and M = 13 is equal to 0.01 and for P = 100 and M = 90, it reaches 0.0004.

### *4.7.7. Flat perfectly conducting rod grating*

The grating is similar to that of figure 4.11, but its height is very small (8 nm). In order to obtain a significant reflection, the dielectric has been replaced by a perfectly conducting material

|  | TE polarization |  | TM polarization |  |
|---|---|---|---|---|
| P,M | $\tau_0$ | $\sum \rho_m + \sum \tau_m$ | $\tau_0$ | $\sum \rho_m + \sum \tau_m$ |
| 100,45 | 0.0598 | 1.1414 | 0.8940 | 0.9759 |
| 200,90 | 0.0634 | 1.0086 | 0.0292 | 0.9887 |
| 300,140 | 0.0639 | 1.0006 | 0.1059 | 0.9996 |
| 400,190 | 0.0639 | 1.0000 | 0.1119 | 1.0000 |

*Table4.7. The flat perfectly conducting rod grating.*

The convergence is very slow and it is necessary to reach P = 300 and M = 140 to obtain a precision better than 0.006, the energy balance being satisfied with a precision better than $10^{-3}$. The reason must be found in the small height of the rods, as explained in section 6.1. The functions included in the kernel of the integral equation become very large in modulus and very small in width for two points located on both sides of the rod for the same abscissa, thus the integration needs a large density of discretization points.

**Appendix 4.A. Mathematical bases of the integral theory**

*4.A.1. Presentation of the mathematical problem*

We consider (figure 4.1) a function $F(x,z) = \begin{cases} F^+(x,z) \text{ in } V^+, \\ F^-(x,z) \text{ in } V^-, \end{cases}$ which satisfies the following conditions:

- it is pseudo-periodic along the x axis:

$$F(x+d,z) = F(x,z)\exp(i\alpha_0 d), \qquad (4.118)$$

- it satisfies a Helmholtz equation:

$$\nabla^2 F^\pm + k^2 \left(n^\pm\right)^2 F^\pm = 0 \quad \text{in } V^\pm, \qquad (4.119)$$

- it satisfies a radiation condition for $z \to \pm\infty$.

The aim of this appendix is to use the second Green's identity and basic theorems on boundary value problems in order to find an integral expression of this function and to analyze the properties of this integral expression. We will deduce the basic keys for writing an integral equation from a boundary value problem.

*4.A.2. Calculation of the Green function*

The first step of the calculation is to find the pseudo-periodic elementary solutions $\mathcal{G}^+(\vec{r})$ and $\mathcal{G}^-(\vec{r})$ of the two Helmholtz equations condensed in eq. (4.119) (with constants $k^2(n^+)^2$ for $\mathcal{G}^+$ and with constant $k^2(n^+)^2$ for $\mathcal{G}^-$, which satisfy the radiation conditions for $z \to \pm\infty$ and the following equations:

$$\nabla^2 \mathcal{G}^+(\vec{r}) + k^2\left(n^+\right)^2 \mathcal{G}^+(\vec{r}) = \sum_{m=-\infty,+\infty} e^{i\alpha_m d}\delta(\vec{r} - md\,\hat{x}), \text{ in } V^+ \text{ and } V^-, \qquad (4.120)$$

$$\nabla^2 \mathcal{G}^-(\vec{r}) + k^2\left(n^-\right)^2 \mathcal{G}^-(\vec{r}) = \sum_{m=-\infty,+\infty} e^{i\alpha_m d}\delta(\vec{r} - md\,\hat{x}), \text{ in } V^+ \text{ and } V^-, \qquad (4.121)$$

$$\mathcal{G}^\pm(\vec{r} + d\,\hat{x}) = \mathcal{G}^\pm(\vec{r})e^{i\alpha_0 d}, \qquad (4.122)$$

with:

$$\alpha_m = \alpha_0 + mK, \qquad K = \frac{2\pi}{d}, \qquad (4.123)$$

and $\hat{x}$ being the unit vector of the x axis.

We must emphasize that, in contrast with eq. (4.119), in which $F^+$ and $F^-$ do not satisfy the same Helmholtz equation, each Green function satisfies a unique Helmholtz equation in the entire space, with constant $k^2(n^+)^2$ for $\mathcal{G}^+$ and with constant $k^2(n^+)^2$ for $\mathcal{G}^-$.

After expanding the periodic function $\mathcal{G}^{\pm} e^{-i\alpha_0 x}$ in Fourier series, then multiplying the Fourier series by $e^{i\alpha_0 x}$, it can be deduced that:

$$\mathcal{G}^{\pm}(x,z) = \sum_{m=-\infty,+\infty} G^{\pm}_m(z) e^{i\alpha_m x}. \tag{4.124}$$

On the other hand, the right-hand member of eq. (4.120), called Dirac comb, can also be expanded in series:

$$\sum_{m=-\infty,+\infty} e^{i\alpha_m d}\delta(\vec{r}-nd\hat{x}) = \frac{1}{d}\delta(z)\sum_m e^{i\alpha_m x}. \tag{4.125}$$

Introducing equations (4.124) and (4.125) in equation (4.120), multiplying by $e^{-i\alpha_0 x}$ then identifying the coefficients of the Fourier series yield:

$$G^{\pm}_m{}''(z) + \left(\gamma^{\pm}_m\right)^2 G^{\pm}_m(z) = \frac{1}{d}\delta(z), \tag{4.126}$$

with $\left(\gamma^{\pm}_m\right)^2 = k^2\left(n^{\pm}\right)^2 - \alpha_m^2$.

For $z \neq 0$, eq.(4.126) becomes the well-known one-dimension propagation equation in a homogeneous media without sources and have for solutions exponentials. We are searching for plane waves that satisfy the radiation condition for $z \rightarrow \pm\infty$, thus:

$$G^{\pm}_m(z) = \begin{cases} A^{\pm}_m \exp\left[i\gamma^{\pm}_m z\right], & \text{if } z > 0, \\ B^{\pm}_m \exp\left[-i\gamma^{\pm}_m z\right], & \text{if } z < 0. \end{cases} \tag{4.127}$$

Distribution theory proves that the solution $G^{\pm}_m$ of eq.(4.126) is a continuous function of z and that its derivative has a jump equal to 1/d at $z=0$. These two conditions allow one to find the unknown amplitudes:

$$\begin{aligned} A^{\pm}_m &= B^{\pm}_m, \\ i\gamma^{\pm}_m A^{\pm}_m - \left(-i\gamma^{\pm}_m B^{\pm}_m\right) &= \frac{1}{d}, \end{aligned} \tag{4.128}$$

and thus:

$$G^{\pm}_m(z) = \frac{1}{2i\gamma^{\pm}_m d}\exp\left(i\gamma^{\pm}_m |z|\right). \tag{4.129}$$

The final form of $\mathcal{G}^{\pm}$ becomes:

$$\mathcal{G}^{\pm}(\vec{r}) = \frac{1}{2id}\sum_m \frac{1}{\gamma^{\pm}_m}\exp\left[i\alpha_m x + i\gamma^{\pm}_m |z|\right]. \tag{4.130}$$

Once the source is not in the origin, the Green functions is the function $\mathcal{G}^{\pm}(\vec{r}-\vec{r}\,')$. Notice that $\mathcal{G}^{\pm}$ symbolizes the Green functions.

*4.A.3. Integral expression*

Now, we apply the second Green theorem in order to find the expression of the function $F^{\pm}$. First, we consider the expression of $F^-$ in $V^-$:

$$F^-(x,z) = \int_{S_T^-} \mathcal{G}^-(x-x',z-z') \frac{dF^-(x',z')}{dN_S} ds' - \int_{S_T^-} \frac{d\mathcal{G}^-(x-x',z-z')}{dN_S} F^-(x',z') ds', \qquad (4.131)$$

with the normal $\vec{N}_S$ being oriented towards the exterior of $V^-$, $x = x(s)$, $x' = x(s')$, and similar expressions for z and z'. The curve $S_T$ includes four parts: the vertical lines $S_L$ at x = 0 and $S_R$ at x = d, the horizontal segment $S_H$ at $z = z_H < 0$ (figure 4.12), and, finally, one period of S. The variable s' denotes the curvilinear abscissa on $S_T^-$, with origine being located at the origin of the Cartesian coordinates.

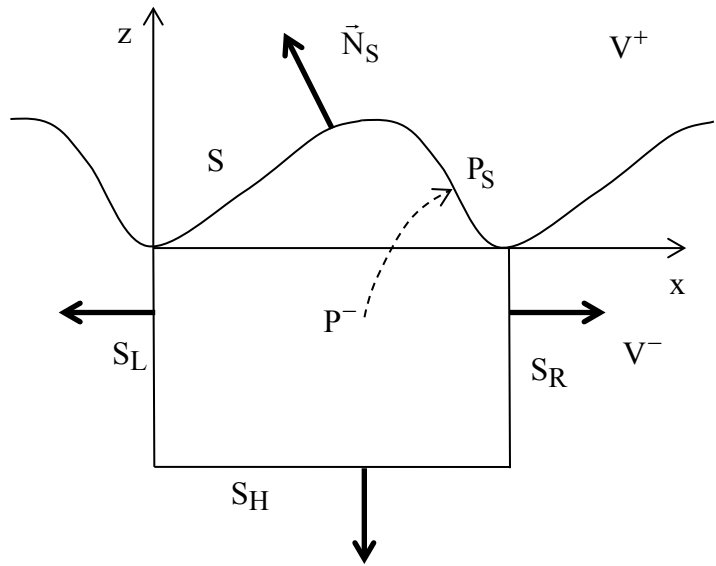

*Figure 4.12. Application of the second Green theorem.*

The pseudo-periodicity in x' of $F^-(x',z')$ (and of $dF^-(x',z')/dN_S$) is opposite to that of $\mathcal{G}^-(x-x',z-z')$ (or of $d\mathcal{G}^-(x-x',z-z')/dN_S$), which entails that $\mathcal{G}^-(x-x',z-z')\frac{dF^-(x',z')}{dN_S}$ and $\frac{d\mathcal{G}^-(x-x',z-z')}{dN_S}F^-(x',z')$ are periodic. Furthermore, taking into account the orientation of the normal on $S_R$ and $S_L$, $\frac{dF^-(x',z')}{dN_S} = -\frac{dF^-(x',z')}{dx}$ and $\frac{d\mathcal{G}^-(x-x',z-z')}{dN_S} = -\frac{d\mathcal{G}^-(x-x',z-z')}{dx}$ on $S_L$, while $\frac{dF^-(x',z')}{dN_S} = +\frac{dF^-(x',z')}{dx}$ on $S_R$. Thus, the integrals on $S_R$ and $S_L$ in eq. (4.131) cancel each other. On $S_H$, ds' and dx' identify and the integral takes the form:

$$\int_{S_H} \left[ \mathcal{G}^-(x-x',z-z')\frac{dF^-(x',z')}{dN_S} - \frac{d\mathcal{G}^-(x-x',z-z')}{dN_S}F^-(x',z') \right] ds'$$

$$= \frac{1}{2d}\sum_{m=-\infty}^{\infty} e^{i\gamma_m^-(z-z_H)} \int_0^d e^{imK(x-x')}\left[\frac{1}{i\gamma_m^-}\phi_H^-(s') + \psi_H^-(s')\right]dx' \quad (4.132)$$

$$= \frac{1}{2}\sum_{m=-\infty}^{\infty} e^{i\gamma_m^-(z-z_H)+imKx}\left(\frac{1}{i\gamma_m^-}\phi_{H,m}^- + \psi_{H,m}^-\right),$$

where $\phi_{H,m}^-$ and $\psi_{H,m}^-$ are the $m^{th}$ Fourier components of the periodic functions $\phi_H^-(s') = \frac{dF^-(x',z')}{dN_S}e^{-i\alpha_0 x'}$ and $\psi_H^-(s') = F^-(x',z')e^{-i\alpha_0 x'}$ defined on $S_H$. Here, we can take advantage of the fact that the segment is parallel to the x axis and lies outside the groove region, which entails that $\phi_{H,m}^-$ and $\psi_{H,m}^-$ are related through the plane wave expansion valid in the homogeneous region (see chapter 2):

$$F^- = \sum_{m=-\infty,+\infty} t_m e^{i\alpha_m x - i\gamma_m^- z} \quad \text{if } z<0, \quad (4.133)$$

which yields:

$$\psi_{H,m}^- = -i\gamma_m^- \phi_{H,m}^-. \quad (4.134)$$

This relation allows us to cancel the integral along $S_H$ and thus the values of $F^-(x,z)$ can be determined by an integral along a single groove of S:

$$F^-(x,z) = \int_{s'=0}^{s_d} \mathcal{G}^-(x-x',z-z')\frac{dF^-(x',z')}{dN_S}ds' - \int_{s'=0}^{s_d} \frac{d\mathcal{G}^-(x-x',z-z')}{dN_S}F^-(x',z')ds', \quad (4.135)$$

with $s_d$ being the curvilinear abscissa of the point of S of abscissa d, the origin of curvilinear abscissa being the origin of the Cartesian coordinates system.

Similar considerations apply to $F^+(x,z)$, so that, after elementary calculations:

$$F^\pm(x,z) = \pm \int_{s'=0}^{s_d} \left[ \mathcal{G}^\pm(x,z,s')e^{i\alpha_0(x-x')}\phi^\pm(s')e^{i\alpha_0 x'} + \mathcal{N}^\pm(x,z,s')e^{i\alpha_0(x-x')}\psi^\pm(s')e^{i\alpha_0 x'} \right]ds' \quad (4.136)$$

with:

$$\mathcal{G}^\pm(x,z,s') = \mathcal{G}^\pm(x,z,x(s'),z(s')) = \frac{1}{2id}\sum_{m=-\infty}^{\infty}\frac{1}{\gamma_m^\pm}e^{imK(x-x'(s'))+i\gamma_m^\pm|z-z'(s')|}, \quad (4.137)$$

$$\mathcal{N}^{\pm}(x,z,s') = -\frac{\partial \mathcal{G}^{\pm}}{\partial N_s}(x,z,x'(s'),z'(s')) =$$

$$= \frac{1}{2d} \sum_{m=-\infty}^{\infty} \left[ \frac{dx'}{ds'} \text{sgn}(z-z') - \frac{\alpha_m}{\gamma_m^{\pm}} \frac{dz'}{ds'} \right] e^{imK(x-x') + i\gamma_m^{\pm}|z-z'|}, \quad (4.138)$$

with $\psi^{\pm}(s') = F^-(x',z')e^{-i\alpha_0 x'}$ and $\phi^{\pm}(s') = \frac{dF^-(x',z')}{dN_s} e^{-i\alpha_0 x'}$ being defined on the grating profile S. Defining the periodic function $U^{\pm}(x,z) = F^{\pm}(x,z)e^{-i\alpha_0 x}$ yields:

$$U^{\pm}(x,z) = \pm \int_{s'=0}^{s_d} \left[ \mathcal{G}^{\pm}(x,z,s')\phi^{\pm}(s') + \mathcal{N}^{\pm}(x,z,s')\psi^{\pm}(s') \right] ds'. \quad (4.139)$$

### *4.A.4. Equation of compatibility*

In this section, we establish a crucial property of the integral theory of gratings, which unfortunately is ignored in most of the reference books of Electromagnetics. With this aim, it is necessary to point out a fundamental property of the expression of $U^{\pm}(x,z)$ given by eqs. (4.13) and (4.139). Let us suppose that we introduce in eq. (4.136) arbitrary periodic functions $\tilde{\phi}^-(s')$ and $\tilde{\psi}^-(s')$. Since we have not introduced the actual physical values $\phi^-(s')$ and $\psi^-(s')$ of these functions, we cannot expect to obtain the actual value of $F^-(x,z)$, but another function $\tilde{F}^-(x,z)$. More important, if the point $P^-(x,z)$ of $V^-$ tends towards a point $P_S$ of curvilinear abscissa s located on the profile (figure 4.12), the limit of $\tilde{F}^-(x,z)$ below the profile, denoted $\lim_{-}\{\tilde{F}^-(x,z)\}$ is not equal to $\tilde{\psi}^-(s')e^{i\alpha_0 x'}$. The same remark can be made for the normal derivative $\lim_{-}\left\{\frac{d\tilde{F}^-(x,z)}{dN_s}\right\}$, which is different from $\tilde{\phi}^-(s')e^{i\alpha_0 x'}$. In order to understand this surprising property, it is necessary to give two results which can be demonstrated using the theory of distributions. Let us give these two fundamental results without demonstration.

    1. The integral expression of $\tilde{F}^-(x,z)$ inside $V^-$ satisfies the Helmholtz equation in $V^-$ like the actual field. What happens to the same integral expression of $\tilde{F}^-(x,z)$, but now calculated in region $V^+$? Denoting by $\tilde{F}^-_{ext}(x,z)$ the function equal to $\tilde{F}^-(x,z)$ in $V^-$ and to this integral extension in $V^+$, it is easy to verify that $\tilde{F}^-_{ext}(x,z)$ satisfies in the entire space the Helmholtz equation satisfied by $\tilde{F}^-(x,z)$ in $V^-$, with constant $k^2(n^-)^2$.

2. The jumps $\lim_+\{\tilde{F}^-_{ext}(x,z)\} - \lim_-\{\tilde{F}^-_{ext}(x,z)\}$ and $\lim_+\left\{\dfrac{d\tilde{F}^-_{ext}(x,z)}{dN_s}\right\} - \lim_-\left\{\dfrac{d\tilde{F}^-_{ext}(x,z)}{dN_s}\right\}$ of $\tilde{F}^-_{ext}(x,z)$ and of its normal derivative across S (difference between the values in $V^+$ and in $V^-$) are respectively equal to $-\tilde{\psi}^-(s')e^{i\alpha_0 x'}$ and $-\tilde{\phi}^-(s')e^{i\alpha_0 x'}$.

The conclusion to draw from these properties is that the limit of $\tilde{F}^-(x,z)$ and of its normal derivative on S are equal to $\tilde{\psi}^-(s')e^{i\alpha_0 x'}$ and $\tilde{\phi}^-(s')e^{i\alpha_0 x'}$ in one case only: if $\tilde{F}^-_{ext}(x,z)$ vanishes throughout $V^+$. Indeed, if this property is satisfied, the jumps are nothing but $-\lim_-\{\tilde{F}^-_{ext}(x,z)\}$ and $-\lim_-\left\{\dfrac{d\tilde{F}^-_{ext}(x,z)}{dN_s}\right\}$, thus $\lim_-\{\tilde{F}^-_{ext}(x,z)\} = \tilde{\psi}^-(s)e^{i\alpha_0 x}$ and $\lim_-\left\{\dfrac{d\tilde{F}^-_{ext}(x,z)}{dN_s}\right\} = \tilde{\phi}^-(s)e^{i\alpha_0 x}$. This case occurs if the values of $\tilde{\psi}^-(s')$ and $\tilde{\phi}^-(s')$ introduced in the integral expression are equal to the actual physical values $\psi^-(s')$ and $\phi^-(s')$.

This result is not surprising for the specialist of boundary value problems. Indeed, using the second Green theorem in $V^-$, we introduce in the integral expression of the field the limit values of both the field and of its normal derivative below S. In the domain of boundary value problems, it is well known that *if a function must satisfy a Helmholtz equation in $V^-$ and a radiation condition for $z \to -\infty$, one cannot impose the limit values of both this function and its normal derivative on S. In fact, we can impose either the limit values of this function or that of its normal derivative on S: in both cases, the solution of the boundary value problem exists and is unique.* Unfortunately, it does not exist any tool of applied mathematics which enables one to express the field in an integral form including either the limit values of the field, or that of its normal derivative on S: a prior solution of an integral equation is required, which is much more difficult.

On the other hand, if both the actual values of the field and of its normal derivative are known, the field can directly be expressed in an integral form through the second Green theorem, without any integral equation, and this is why the second Green theorem is considered as a basic tool of the integral mathods of scattering. It must be emphasized that in that case, we know *a priori* that the the limit values of the field and of its normal derivative are the actual ones, thus that they are compatible, which is not the case for arbitrarily chosen limits.

This fundamental property shows that when the second Green theorem is used with unknown values of the limits of the field and of its normal derivative, we must impose to these limits an equation of compatibility. This compatibility is satisfied if we impose to the limit $\lim_-\{\tilde{F}^-_{ext}(x,z)\}$ of the integral expression of $\tilde{F}^-_{ext}(x,z)$ to be equal to $\tilde{\psi}^-(s)e^{i\alpha_0 x}$ or if we impose to the limit $\lim_+\{\tilde{F}^-_{ext}(x,z)\}$ of the integral expression of $\tilde{F}^-_{ext}(x,z)$ to be equal to zero, these two conditions being equivalent. Indeed, if $\lim_+\{\tilde{F}^-_{ext}(x,z)\} = 0$, the expression of $\tilde{F}^-_{ext}(x,z)$ in $V^+$ satisfies a Helmholtz equation, a radiation condition at infinity and its limit on S vanishes. The obvious solution of this boundary value problem is $\tilde{F}^-_{ext}(x,z) = 0$ in $V^+$, and we have seen that the solution of this boundary value problem is unique, thus it is the solution. The consequence is that $\lim_+ \left\{ \dfrac{d\tilde{F}^-_{ext}(x,z)}{dN_S} \right\} = 0$, thus $\lim_- \left\{ \dfrac{d\tilde{F}^-_{ext}(x,z)}{dN_S} \right\} = \tilde{\phi}^-(s)e^{i\alpha_0 x}$.

In order to implement this condition of compatibility, we have to express $\lim_+ \left\{ \dfrac{d\tilde{F}^-_{ext}(x,z)}{dN_S} \right\}$ or $\lim_- \left\{ \dfrac{d\tilde{F}^-_{ext}(x,z)}{dN_S} \right\}$. We can use a first equation:

$$\lim_+ \left\{ \tilde{F}^-_{ext}(x,z) \right\} - \lim_- \left\{ \tilde{F}^-_{ext}(x,z) \right\} = -\tilde{\psi}^-(s)e^{i\alpha_0 x}, \qquad (4.140)$$

or equivalently:

$$\lim_+ \left\{ \tilde{U}^-_{ext}(x,z) \right\} - \lim_- \left\{ \tilde{U}^-_{ext}(x,z) \right\} = -\tilde{\psi}^-(s). \qquad (4.141)$$

In order to find a second equation, we can consider eq. (4.139) which gives the expression of $U^-(x,z) = F^-(x,z)e^{-i\alpha_0 x}$. If z is fixed, this expression is a Fourier series in x, which is discontinuous on S. It is well known that the value on S of this Fourier series is the average value of the limits on both sides of S, thus:

$$\lim_+ \left\{ \tilde{U}^-_{ext}(x,z) \right\} + \lim_- \left\{ \tilde{U}^-_{ext}(x,z) \right\} = -2 \int_{s'=0}^{s_d} \left[ \mathcal{G}^-(s,s')\tilde{\phi}^-(s') + \mathcal{N}^-(s,s')\tilde{\psi}^-(s') \right] ds', \qquad (4.142)$$

with $\mathcal{G}^-(s,s')$ and $\mathcal{N}^-(s,s')$ being the integral expressions of $\mathcal{G}^-(x,z,s')$ and $\mathcal{N}^-(x,z,s')$ when the point of coordinates x,z becomes a point of S of curvilinear abscissa s:

$$\mathcal{G}^-(s,s') = \mathcal{G}^-(x(s),z(s),s'), \qquad (4.143)$$

$$\mathcal{N}^-(s,s') = \mathcal{N}^-(x(s),z(s),s'). \qquad (4.144)$$

From eqs. (4.141) and (4.142), we deduce the limits of $\tilde{U}^-$ on both sides of S, and we derive the equation of compatibility in $V^-$:

$$\int_{s'=0}^{s_d} \left[ \mathcal{G}^-(s,s')\tilde{\phi}^-(s') + \mathcal{N}^-(s,s')\tilde{\psi}^-(s') \right] ds' + \dfrac{\tilde{\psi}^-(s)}{2} = 0. \qquad (4.145)$$

Achieving a similar calculation for $V^+$ yields the general equation of compatibility in $V^\pm$:

$$\int_{s'=0}^{s_d} \left[ \mathcal{G}^\pm(s,s')\tilde{\phi}^\pm(s') + \mathcal{N}^\pm(s,s')\tilde{\psi}^\pm(s') \right] ds' \mp \dfrac{\tilde{\psi}^\pm(s)}{2} = 0, \qquad (4.146)$$

$$\mathcal{G}^\pm(s,s') = \dfrac{1}{2id} \sum_{m=-\infty}^{\infty} \dfrac{1}{\gamma^\pm_m} e^{imK(x(s)-x'(s')) + i\gamma^\pm_m |z(s)-z'(s')|}, \qquad (4.147)$$

$$\mathcal{N}^{\pm}(s,s') = \frac{1}{2d} \sum_{m=-\infty}^{\infty} \left[ \frac{dx'}{ds'} \text{sgn}(z(s)-z'(s')) - \frac{\alpha_m}{\gamma_m^{\pm}} \frac{dz'}{ds'} \right] e^{imK(x(s)-x'(s'))+i\gamma_m^{\pm}|z(s)-z'(s')|}. \quad (4.148)$$

It can be shown [7] that $\mathcal{G}^{\pm}(s,s')$ has an integrable logarithmic singularity (it behaves like $a_0 + a_1 \text{Log}|s-s'|$ when $s' \to s$, $a_0$ and $a_1$ complex numbers) which can be taken into account in the integral by removing the singularity $a_1 \text{Log}|s-s'|$ from $\mathcal{G}^{\pm}(s,s')$ then by integrating it in closed form. At the first glance, $\mathcal{N}^{\pm}(s,s')$ cannot be continuous, due to the discontinuity of $\text{sgn}(z(s)-z'(s'))$ for s=s'. In fact, a careful analysis of this function around s = s' shows that it is continuous and that its limit when $s' \to s$ can be expressed in closed form [44,7], as stated in section 6.3.

### 4.A.5. Generalized compatibility

In the physical problem, the total field in $V^+$ includes the incident field. Here, we give an extension of the compatibility equation to the total field which allows a significant simplification of the use of integral theory.

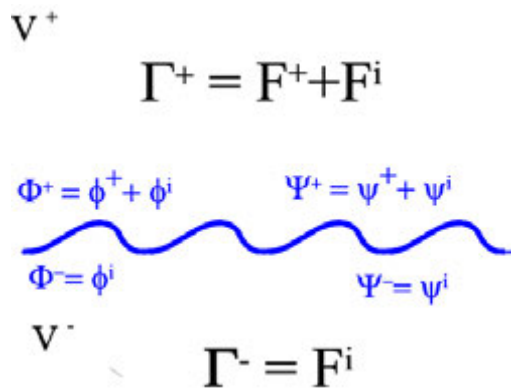

*Figure 4.13. generalized: generalized compatibility*

We define (figure 4.13) a function $\Gamma = \begin{cases} \Gamma^+ = F^{T+} = F^+ + F^i & \text{in } V^+, \\ \Gamma^- = F^i & \text{in } V^-. \end{cases}$

$F^+$ being the field scattered in $V^+$ by a grating illuminated by the incident field $F^i$. $\Gamma^-$ is the expression of the incident field in $V^-$. Thus, in contrast with the preceding section, the function considered in this section does not satisfy a radiation condition at infinity in $V^+$. According to eq. (4.136), the value of $F^+$ is given by:

$$F^+(x,z) = \int_{s'=0}^{S_d} \left[ \mathcal{G}^+(x,z,s')e^{i\alpha_0(x-x')}\phi^+(s')e^{i\alpha_0 x'} \right.$$
$$\left. + \mathcal{N}^+(x,z,s')e^{i\alpha_0(x-x')}\psi^+(s')e^{i\alpha_0 x'} \right] ds', \quad (4.149)$$

thus $\Gamma^+$ can be written:

$$\Gamma^+(x,z) = F^i(x,y) + \int_{s'=0}^{S_d} \left[ \mathcal{G}^+(x,z,s') e^{i\alpha_0(x-x')} \phi^+(s') e^{i\alpha_0 x'} \right. \tag{4.150}$$
$$\left. + \mathcal{N}^+(x,z,s') e^{i\alpha_0(x-x')} \psi^+(s') e^{i\alpha_0 x'} \right] ds'.$$

We denote by $\Psi^+ = \psi^+ + \psi^i$ and $\Phi^+ = \phi^+ + \phi^i$ the limits of $\Gamma^+$ on S and its normal derivative respectively. Introducing these values in eq. (4.149) yields:

$$F^+(x,z) = \int_{s'=0}^{S_d} \left[ \mathcal{G}^+(x,z,s') e^{i\alpha_0(x-x')} \left(\Phi^+(s') - \phi^i(s')\right) e^{i\alpha_0 x'} \right. \tag{4.151}$$
$$\left. + \mathcal{N}^+(x,z,s') e^{i\alpha_0(x-x')} \left(\Psi^+(s') - \psi^i(s')\right) e^{i\alpha_0 x'} \right] ds'.$$

Thus, the integral expression of the total field $\Gamma^+$ is given by:

$$\Gamma^+(x,z) = F^i(x,z) + \int_{s'=0}^{S_d} \left[ \mathcal{G}^+(x,z,s') e^{i\alpha_0(x-x')} \left(\Phi^+(s') - \phi^i(s')\right) e^{i\alpha_0 x'} \right.$$
$$\left. + \mathcal{N}^+(x,z,s') e^{i\alpha_0(x-x')} \left(\Psi^+(s') - \psi^i(s')\right) e^{i\alpha_0 x'} \right] ds', \tag{4.152}$$

or, gathering the terms containing the incident field:

$$\Gamma^+(x,z) =$$
$$F^i(x,z) - \int_{s'=0}^{S_d} \left[ \mathcal{G}^+(x,z,s') e^{i\alpha_0(x-x')} \phi^i(s') e^{i\alpha_0 x'} + \mathcal{N}^+(x,z,s') e^{i\alpha_0(x-x')} \psi^i(s') e^{i\alpha_0 x'} \right] ds' \tag{4.153}$$
$$+ \int_{s'=0}^{S_d} \left[ \mathcal{G}^+(x,z,s') e^{i\alpha_0(x-x')} \Phi^+(s') e^{i\alpha_0 x'} + \mathcal{N}^+(x,z,s') e^{i\alpha_0(x-x')} \Psi^+(s') e^{i\alpha_0 x'} \right] ds'.$$

In order to simplify this equation, we consider the part of the integral containing the incident field, the middle line in the equation above. This part is equal to 0. Indeed, we know that $\phi^i(s')$ and $\psi^i(s')$ are compatible in $V^-$ since they are derived from the actual values of the incident field in $V^-$ (figure 4.13). We have shown in the preceding section that the integral expression of such a function vanishes in $V^+$. Remarking that $\mathcal{G}^-(x,z,s') = \mathcal{G}^+(x,z,s')$ and $\mathcal{N}^-(x,z,s') = \mathcal{N}^+(x,z,s')$ since $\Gamma^+$ and $\Gamma^-$ satisfy the same Helmholtz equation (with constant $k^2(n^+)^2$), we can write than in $V^+$:

$$\int_{s'=0}^{S_d} \left[ \mathcal{G}^+(x,z,s') e^{i\alpha_0(x-x')} \phi^i(s') e^{i\alpha_0 x'} + \mathcal{N}^+(x,z,s') e^{i\alpha_0(x-x')} \psi^i(s') e^{i\alpha_0 x'} \right] ds' = 0. \tag{4.154}$$

Thus finally the expression of $\Gamma^+$ is given by:

$$\Gamma^+(x,z) = F^i(x,z) + \int_{s'=0}^{s_d} \left[ \mathcal{G}^+(x,z,s')e^{i\alpha_0(x-x')}\Phi^+(s')e^{i\alpha_0 x'} \right.$$
$$\left. + \mathcal{N}^+(x,z,s')e^{i\alpha_0(x-x')}\Psi^+(s')e^{i\alpha_0 x'} \right]ds'. \quad (4.155)$$

*The result is that the expression of the **total** field in $V^+$ can be obtained from its limit value $\Psi^+$ on S and from its normal derivative $\Phi^+$, by adding the incident field to the integral expression deduced from the Green theorem.*

The generalized compatibility equation is derived from the compatibility equations for $F^+$ in $V^+$ and for $F^i$ in $V^-$:

$$\int_{s'=0}^{s_d} \left[ \mathcal{G}^+(s,s')\phi^+(s') + \mathcal{N}^+(s,s')\psi^+(s') \right]ds' - \frac{\psi^+(s)}{2} = 0, \quad (4.156)$$

$$\int_{s'=0}^{s_d} \left[ \mathcal{G}^+(s,s')\phi^i(s') + \mathcal{N}^+(s,s')\psi^i(s') \right]ds' + \frac{\psi^i(s)}{2} = 0. \quad (4.157)$$

By adding these two equations, we deduce that:

$$\int_{s'=0}^{s_d} \left[ \mathcal{G}^+(s,s')\Phi^+(s') + \mathcal{N}^+(s,s')\Psi^+(s') \right]ds' + \psi^i = \frac{\Psi^+(s)}{2}. \quad (4.158)$$

The remarkable result is that the compatibility equation given in the preceding section can be extended to the total field: the left-hand side of eq. (4.158) represents the expression of the total field on S and the right-hand, one half of its limit on S. Thus the generalized compatibility condition can be stated in the following way:

The compatibility condition for a field in $V^+$ including incident and/or diffracted waves and satisfying the two conditions:

- it is pseudo-periodic along the x axis:

$$F(x+d,z) = F(x,z)\exp(i\alpha_0 d), \quad (4.159)$$

- it satisfies a Helmholtz equation:

$$\nabla^2 F^\pm + k^2(n^\pm)^2 F^\pm = 0 \quad \text{in } V^\pm, \quad (4.160)$$

can be stated in the following way: *The value on S of the **total** field, obtained by adding the incident wave to the integral expression deduced from the second Green theorem (but in which the limits are those of the total field) is equal to the half of its limit value on S.*

A similar generalized compatibility condition can be obtained for a total field in $V^-$ when an incident wave propagating upward in $V^-$ (supposed to contain a lossless dielectric) illuminates the grating surface, but this case is not worth in the frame of this chapter.

## 4.A.6. Normal derivative of a field continuous on S.

The calculation of the normal derivative of $F^\pm$ on the grating surface in the general case is difficult. However, this aim can be reached at least in one case: when it is possible to define both $F^+$ and $F^-$ which satisfy three conditions:

- F is continuous across S, or equivalently $\psi^+(s) = \psi^-(s)$,
- $F^+$ and $F^-$ satisfy the same Helmholtz equation, with constant $k^2(n^+)^2$, or equivalently $n^+ = n^-$.
- F satisfies a radiation condition at infinity.

Due to the second condition, it seems that this case does not make sense: if $n^+ = n^-$, one cannot expect any scattering phenomenon. However, the study of this purely mathematical problem is crucial, for example in the study of perfectly conducting gratings.

First, it is worth noting that, thanks to the first condition, the gradient of F can be calculated without any use of distributions, which is not the case if F is discontinuous on S. Since $\gamma_m^- = \gamma_m^+$, $\mathcal{G}^-(x,z,s') = \mathcal{G}^+(x,z,s')$ and $\mathcal{N}^-(x,z,s') = \mathcal{N}^+(x,z,s')$ it can be deduced from eq. (4.139) that:

$$F^+(x,z) = \int_{s'=0}^{S_d} \left[ \mathcal{G}^+(x,z,s')e^{i\alpha_0 x}\phi^+(s') + \mathcal{N}^+(x,z,s')e^{i\alpha_0 x}\psi^+(s') \right] ds', \qquad (4.161)$$

$$F^-(x,z) = -\int_{s'=0}^{S_d} \left[ \mathcal{G}^+(x,z,s')e^{i\alpha_0 x}\phi^-(s') + \mathcal{N}^+(x,z,s')e^{i\alpha_0 x}\psi^-(s') \right] ds'. \qquad (4.162)$$

We have seen in this appendix that, if $\phi^+(s')$ and $\psi^+(s')$ are compatible, the expression of $F^+(x,z)$ given by eq. (4.161) vanishes in $V^-$. The same property holds for the expression of $F^-(x,z)$ given by eq. (4.162), which vanishes in $V^+$ if $\phi^-(s')$ and $\psi^-(s')$ are compatible. Bearing in mind that $\psi^+(s) = \psi^-(s)$, the expression of F in the entire space is given by adding the right-hand sides of eqs. (4.161) and (4.162):

$$F(x,z) = F^+(x,z) + F^-(x,z) = \int_{s'=0}^{S_d} \mathcal{G}^+(x,z,s')e^{i\alpha_0 x}\left(\phi^+(s') - \phi^-(s')\right) ds'. \qquad (4.163)$$

Thanks to the continuity of F on S, the expression of its value on the profile does not make problem and is given by:

$$F(s) = \int_{s'=0}^{S_d} \mathcal{G}^+(s,s')e^{i\alpha_0 x}\left(\phi^+(s') - \phi^-(s')\right) ds'. \qquad (4.164)$$

In order to obtain the normal derivative of F, let us calculate its gradient:

$$\nabla F(x,z) = \int_{s'=0}^{S_d} \nabla_{(x,z)} \left[ \mathcal{G}^+(x,z,s')e^{i\alpha_0 x} \right] \left( \phi^+(s') - \phi^-(s') \right) ds', \qquad (4.165)$$

$$\nabla \left[ \mathcal{G}^+(x,z,s')e^{i\alpha_0 x} \right] = \frac{1}{2d} \begin{pmatrix} \sum_{m=-\infty,+\infty} \frac{\alpha_m}{\gamma_m^+} e^{i\alpha_0 x} e^{imK(x-x')+i\gamma_m^+|z-z'|} \\ \sum_{m=-\infty,+\infty} e^{i\alpha_0 x} e^{imK(x-x')+i\gamma_m^+|z-z'|} \end{pmatrix}. \qquad (4.166)$$

The components of the normal $\vec{N}_S$ are given by:

$$\vec{N}_S = \begin{pmatrix} -\dfrac{dy}{ds} \\ \dfrac{dx}{ds} \end{pmatrix}, \qquad (4.167)$$

and thus:

$$\left[ \frac{d\mathcal{G}^+(x,z,s')e^{i\alpha_0 x}}{dN_S} \right]^{\pm} = \qquad (4.168)$$

$$\frac{1}{2d} \lim_{\pm} \left\{ \sum_{m=-\infty,+\infty} \left[ \text{sgn}(z-z')\frac{dx}{ds} - \frac{\alpha_m}{\gamma_m^+}\frac{dy}{ds} \right] e^{i\alpha_0 x} e^{imK(x-x')+i\gamma_m^+|z-z'|} \right\}.$$

Using eqs. (4.165) and (4.168) yields:

$$\frac{dF^{\pm}}{dN_S} = \int_{s'=0}^{S_d} \left[ \frac{d\mathcal{G}^+(x,z,s')e^{i\alpha_0 x}}{dN_S} \right]^{\pm} \left( \phi^+(s') - \phi^-(s') \right) ds'. \qquad (4.169)$$

In order to eliminate the use of limits in the expression of $\left[ \dfrac{d\mathcal{G}^+(x,z,s')e^{i\alpha_0 x}}{dN_S} \right]^{\pm}$ given by eq. (4.168), it can be remembered that, by definition,

$$\frac{dF^+}{dN_S} - \frac{dF^-}{dN_S} = \left[ \phi^+(s) - \phi^-(s) \right] e^{i\alpha_0 x}. \qquad (4.170)$$

Moreover, it is to be noticed that, when z is constant, the components of $\nabla \left[ \mathcal{G}^+(x,z,s')e^{i\alpha_0 x} \right]$ given by eq. (4.166) are Fourier series in x. Using again the property of discontinuous Fourier series on the discontinuity, it can be derived that:

$$\frac{dF^+}{dN_S} + \frac{dF^-}{dN_S} =$$

$$\frac{1}{d} \int_{s'=0}^{S_d} \sum_{m=-\infty}^{+\infty} \left[ \text{sgn}(z-z')\frac{dx}{ds} - \frac{\alpha_m}{\gamma_m^+}\frac{dy}{ds} \right] e^{i\alpha_0 x} e^{imK(x-x')+i\gamma_m^+|z-z'|} \left[ \phi^+(s') - \phi^-(s') \right] ds'. \quad (4.171)$$

From equations (4.170) and (4.171), we deduce:

$$\frac{dF^\pm}{dN_S} = \pm \frac{\phi^+(s) - \phi^-(s)}{2} e^{i\alpha_0 x} + \int_{s'=0}^{S_d} \mathcal{K}(s,s') e^{i\alpha_0 x} \left[ \phi^+(s') - \phi^-(s') \right] ds', \quad (4.172)$$

with:

$$\mathcal{K}(s,s') = \frac{1}{2d} \sum_{m=-\infty,+\infty} \left[ \text{sgn}(z-z')\frac{dx}{ds} - \frac{\alpha_m}{\gamma_m^+}\frac{dy}{ds} \right] e^{imK(x-x')+i\gamma_m^+|z-z'|}. \quad (4.173)$$

It is interesting to notice that the expression of function $\mathcal{K}(s,s')$ is very close to that of $\mathcal{N}(s,s')$, the only difference being that $\frac{dx'}{ds'}$ and $\frac{dy'}{ds'}$ are replaced by $\frac{dx}{ds}$ and $\frac{dy}{ds}$.

Like $\mathcal{N}^\pm(s,s')$, $\mathcal{K}(s,s')$ is continuous and its limit when $s' \to s$ can be expressed in closed form [44,7], as stated in section 6.3.

### *4.A.7. Limit values of a field with continuous normal derivative on S.*

We consider a function F satisfying the following conditions
- F has the same normal derivative on both sides of S, or equivalently $\phi^+(s) = \phi^-(s)$,
- $F^+$ and $F^-$ satisfy the same Helmholtz equation, or equivalently $n^+ = n^-$.
- F satisfies a radiation condition at infinity.

The aim of this section is to calculate the limits of F on both parts of S.
From the secong Green theorem (eq. (4.136)), F can be expressed from the values of $\phi^+(s)$, $\phi^-(s)$, $\psi^+(s)$ and $\psi^-(s)$:

$$F^+(x,z) = \int_{s'=0}^{S_d} \left[ \mathcal{G}^+(x,z,s') e^{i\alpha_0(x-x')} \phi^+(s') e^{i\alpha_0 x'} \right.$$

$$\left. + \mathcal{N}^+(x,z,s') e^{i\alpha_0(x-x')} \psi^+(s') e^{i\alpha_0 x'} \right] ds', \quad (4.174)$$

and, bearing in mind that $\phi^-(s) = \phi^+(s)$ and that $n^- = n^+$, $\mathcal{G}^-(x,z,s') = \mathcal{G}^+(x,z,s')$ and $\mathcal{N}^-(x,z,s') = \mathcal{N}^+(x,z,s')$:

$$F^-(x,z) = -\int_{s'=0}^{s_d} \left[ \mathcal{G}^+(x,z,s')e^{i\alpha_0(x-x')}\phi^+(s')e^{i\alpha_0 x'} \right. \\ \left. + \mathcal{N}^+(x,z,s')e^{i\alpha_0(x-x')}\psi^-(s')e^{i\alpha_0 x'} \right] ds'. \quad (4.175)$$

It has been shown in section 4.A.4 that, if $\phi^+(s')$ and $\psi^+(s')$ are compatible, the expression of $F^+(x,z)$ given by eq. (4.174) vanishes in $V^-$. The same property holds for the expression of $F^-(x,z)$ given by eq. (4.175), which vanishes in $V^+$, thus we can write that, if the compatibility equations are satisfied:

$$F(x,z) = F^+(x,z) + F^-(x,z) = \int_{s'=0}^{s_d} \mathcal{N}^+(x,z,s')e^{i\alpha_0 x}\Psi(s')ds', \quad (4.176)$$

or

$$U(x,z) = F(x,z)e^{-i\alpha_0 x} = \int_{s'=0}^{s_d} \mathcal{N}^+(x,z,s')\Psi(s')ds', \quad (4.177)$$

with:

$$\Psi(s') = \psi^+(s') - \psi^-(s'). \quad (4.178)$$

In order to express the limits $\lim\{U^+(x,z)\}$ or $\lim\{U^-(x,z)\}$ on both parts of S, we can use a first equation:

$$\lim_+\{U^+(x,z)\} - \lim_-\{U^-(x,z)\} = \Psi(s). \quad (4.179)$$

To find a second equation, we can consider eqs. (4.176) and (4.177) which gives the expression of $U(x,z)$. If z is fixed, the expression of $U(x,z)$ is a Fourier series in x, which is discontinuous on S. The value on S of this Fourier series is the average value of the limits on both sides of S, thus:

$$\lim_+\{U^+(x,z)\} + \lim_-\{U^-(x,z)\} = 2\int_{s'=0}^{s_d} \mathcal{N}^+(s,s')\Psi(s')ds'. \quad (4.180)$$

From eqs. (4.179) and (4.180), we deduce the two limits:

$$\lim_\pm\{U(x,y)\} = \pm\frac{\Psi(s)}{2} + \int_{s'=0}^{s_d} \mathcal{N}^+(s,s')\Psi(s')ds'. \quad (4.181)$$

*4.A.8. Calculation of the amplitudes of the plane wave expansions at infinity.*

For $z \to \pm\infty$, the expression of $F^\pm$ given by eq. (4.136) can be simplified since $\text{sgn}(z-z') = \pm 1$ and $|z-z'| = \pm(z-z')$, in such a way that the expression of F at infinity becomes a sum of plane waves:

$$F^+(x,y) = \sum_{m=-\infty}^{\infty} r_m \exp\left(i\alpha_m x + i\gamma_m^+ z\right) \text{ if } z > z_0, \quad (4.182)$$

$$F^-(x,y) = \sum_{m=-\infty}^{\infty} t_m \exp\left(i\alpha_m x - i\gamma_m^- z\right) \text{ if } z < 0, \quad (4.183)$$

$$r_m = \frac{1}{2d} \int_{s=0}^{S_d} e^{-imKx(s)-i\gamma_m^+ z(s)} \left[ \frac{-i\phi^+(s)}{\gamma_m^+} + \left(\frac{dx(s)}{ds} - \frac{\alpha_m}{\gamma_m^+} \frac{dz(s)}{ds}\right) \psi^+(s) \right] ds, \quad (4.184)$$

$$t_m = \frac{1}{2d} \int_{s=0}^{S_d} e^{-imKx(s)+i\gamma_m^- z(s)} \left[ \frac{i\phi^-(s)}{\gamma_m^-} + \left(\frac{dx(s)}{ds} + \frac{\alpha_m}{\gamma_m^+} \frac{dz(s)}{ds}\right) \psi^-(s) \right] ds, \quad (4.185)$$

with $z_0$ being the ordinate of the top of the grating profile. It must be noticed that a finite number of orders m, called propagating orders, are non-evanescent and propagate at infinity. They correspond to real values of $\gamma_m^+$ (for reflected orders) or $\gamma_m^-$ (for transmitted orders, if the optical index $n^-$ is real only).

Equation (4.184) can easily be generalized to the case in which we know the limit value of the total field on S (including incident waves) and its normal derivative. It suffices to analyze the behaviour at infinity of eq. (4.155) instead of eq. (4.136). The result is that it suffices to replace the values $\phi^+(s)$ and $\psi^+(s)$ relative to the scattered field by the values $\Phi^+(s)$ and $\Psi^+(s)$ relative to the total field:

$$r_m = \frac{1}{2d} \int_{s=0}^{S_d} e^{-imKx(s)-i\gamma_m^+ z(s)} \left[ \frac{-i\Phi^+(s)}{\gamma_m^+} + \left(\frac{dx(s)}{ds} - \frac{\alpha_m}{\gamma_m^+} \frac{dz(s)}{ds}\right) \Psi^+(s) \right] ds. \quad (4.186)$$

Diffraction efficiencies $\rho_m$ in the reflected orders propagating above the grating can be obtained by using the Poynting theorem on segments of one period parallel to the x axis:

$$\rho_m = \frac{\gamma_m^+}{\gamma_0^+} |r_m|^2. \quad (4.187)$$

When the grating is made of a lossless dielectric, the transmitted efficiencies $\tau_m$ are given by:

$$\tau_m = \frac{q^-}{q^+} \frac{\gamma_m^+}{\gamma_0^+} |t_m|^2. \quad (4.188)$$

**Appendix 4.B. Integral method leading to a single integral equation for bare, metallic or dielectric grating**

Historically, the formalism presented in this Appendix was the first one to lead to a single integral equation for a dielectric or metallic grating. The steps of the method are summarized in figure 4.14.

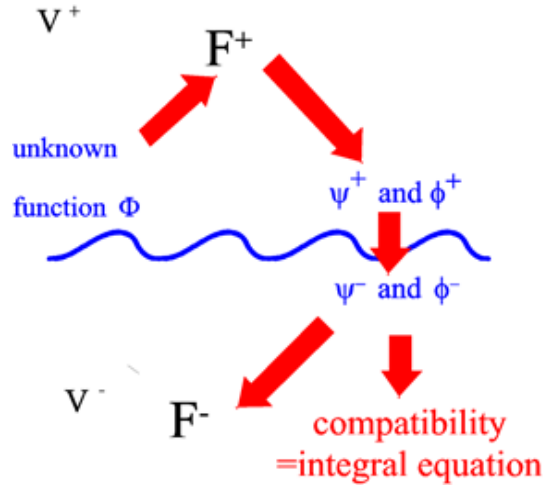

*Figure 4.14. Steps of the integral formalism leading to a single equation*

### *4.B.1. Definition of the unknown function*

The single unknown function $\Phi$ is defined from a function $\tilde{\Gamma} = \begin{cases} \tilde{\Gamma}^+ \text{ in } V^+, \\ \tilde{\Gamma}^- \text{ in } V^-, \end{cases}$ satisfying the following conditions:

- it satisfies the same Helmholtz equation in the entire space, except may be on the profile S,:

$$\nabla^2 \tilde{\Gamma} + k^2 \left(n^+\right)^2 \tilde{\Gamma} = 0, \qquad (4.189)$$

- it has the same pseudo-periodicity as the actual solution of the grating problem:

$$\tilde{\Gamma}(x+d, z) = \tilde{\Gamma}(x, z) e^{i\alpha_0 d}, \qquad (4.190)$$

- it identifies to the actual physical solution of the grating problem in $V^+$:

$$\tilde{\Gamma}^+ \equiv F^+, \qquad (4.191)$$

- it is continuous across S,
- it satisfies a radiation condition for $y \to \pm\infty$.

We denote by $\psi'^{\pm} e^{i\alpha_0 x'}$ and $\phi'^{\pm} e^{i\alpha_0 x'}$ the limit values of $\tilde{\Gamma}$ and of its normal derivative $\dfrac{d\tilde{\Gamma}}{dN_S}$ on S, bearing in mind that by definition, $\psi'^+ \equiv \psi^+$ and $\phi'^+ \equiv \phi^+$.

The question which arises is to know if $\tilde{\Gamma}^-$ is well defined. The continuity of $\tilde{\Gamma}$ across S imposes the value of $\tilde{\Gamma}^-$ on S: $\psi'^- \equiv \psi'^+ \equiv \psi^+$. In addition, $\tilde{\Gamma}^-$ satisfies a Helmholtz equation and a radiation condition at infinity. It is worth noting that, in contrast with the function $\Gamma$ introduced in Appendix 4.A, here the function $\tilde{\Gamma}$ does not include the incident field and thus has no physical meaning below the profile. As mentioned in Appendix 4.A, the solution of this boundary value problem (since we impose that $\tilde{\Gamma}^- = \tilde{\Gamma}^+$) exists and is unique, thus $\tilde{\Gamma}$ is correctly defined in the entire space. The unknown function $\Phi$ of the integral equation is defined as the jump of the normal derivative of $\tilde{\Gamma}$ across S, more precisely:

$$\Phi = \phi'^+ - \phi'^-. \tag{4.192}$$

As regards the physical interpretation of $\Phi$, it can be shown easily that $\Phi \, e^{i\alpha_0 x}$ is the surface current density which, placed on S, generates in $V^+$ the actual scattered field. In other words, we have considered a fictitious structure consisting of an infinitely thin, perfectly conducting metallic sheet supporting a surface current density $j(s)\hat{y}$ placed on the grating surface, this surface separating two media having identical optical properties (refractive index $n^+$). The unknown $\Phi$ is proportional to j.

### 4.B.2. Expression of the scattered field, its limit on S and its normal derivative from $\Phi$.

It must be noticed that the function $\tilde{\Gamma}$ satisfies all the conditions of the function F of section 4.A.5. Since $\tilde{\Gamma}$ is continuous on S, the calculation of $\tilde{\Gamma}$ from $\Phi$ can be achieved using eqs. (4.163) and (4.137):

$$\tilde{\Gamma}(x,z) = \int_{s'=0}^{S_d} \mathcal{G}^+(x,z,s') e^{i\alpha_0 x} \Phi(s') ds', \tag{4.193}$$

$$\mathcal{G}^+(x,y,s') = \frac{1}{2id} \sum_{m=-\infty}^{\infty} \frac{1}{\gamma_m^+} \exp\left[ imK(x-x') + i\gamma_m^+ |z-z'| \right]. \tag{4.194}$$

The value of the limit $\psi'^+(s) e^{i\alpha_0 x}$ of $\tilde{\Gamma}^+(x,z)$ on S does not make problem, thanks to its continuity:

$$\psi'^+(s) = \int_{s'=0}^{S_d} \mathcal{G}^+(s,s') \Phi(s') ds', \tag{4.195}$$

with $\mathcal{G}^+(s,s')$ the value on S of $\mathcal{G}^+(x,y,s')$, given by eq. (4.147):

$$\mathcal{G}^+(s,s') = \frac{1}{2id} \sum_{m=-\infty}^{\infty} \frac{1}{\gamma_m^+} e^{imK(x(s)-x'(s')) + i\gamma_m^+ |z(s)-z'(s')|}, \tag{4.196}$$

and finally its normal derivative can be derived from eq. (4.172) and (4.173):

$$\phi^+(s) = \frac{\Phi(s)}{2} + \int_{s'=0}^{S_d} \mathcal{K}(s,s') \Phi(s') ds', \tag{4.197}$$

$$\mathcal{K}(s,s') = \frac{1}{2d} \sum_{m=-\infty,+\infty} \left[ \text{sgn}(z-z')\frac{dx}{ds} - \frac{\alpha_m}{\gamma_m^+}\frac{dy}{ds} \right] e^{imK(x-x')+i\gamma_m^+|z-z'|}. \qquad (4.198)$$

It is worth noting that the values of the limits of the field and its normal derivative on S that are the two unknown functions in the classical integral theory (section 2), are now expressed from the single function $\Phi(s)$. This is not surprising since the limits on S of the function $\tilde{\Gamma}$ given by eq. (4.193) satisfy automatically a relation of compatibility, whatever the function $\Phi(s')$ introduced in the integral may be: it is the field generated by a surface current on S. As a consequence, we have not to include in the theory a relation of compatibility in $V^+$, which was the first integral equation in the classical formalism.

### 4.B.3. Integral equation

The single integral equation will be obtained by writing the relation of compatibility in $V^-$, considered now to be filled by the actual grating material, i.e. a material of index $n^-$. With this aim, we calculate the limits in $V^-$ of the field and its normal derivative on S. inserting in the continuity conditions of the field given by eqs (4.8) and (4.10) the the limit values given by eqs (4.195) and (4.197):

$$\psi^-(s) = \psi^i(s) + \int_{s'=0}^{S_d} \mathcal{G}^+(s,s')\Phi(s')ds', \qquad (4.199)$$

$$\phi^-(s) = \frac{q^+}{q^-}\left[\frac{\Phi(s)}{2} + \phi^i(s) + \int_{s'=0}^{S_d} \mathcal{K}(s,s')\Phi(s')ds'\right], \qquad (4.200)$$

with $q^\pm$ given by eq. (4.12).

The field in $V^-$ can be deduced from eqs. (4.199) and (4.200) using eq. (4.136):

$$F^-(x,z) = -e^{i\alpha_0 x}\int_{s'=0}^{S_d}\left[\mathcal{G}^-(x,z,s')\phi^-(s') + \mathcal{N}^-(x,z,s')\psi^-(s')\right]ds', \qquad (4.201)$$

and the equation of compatibility is given by eq. (4.146):

$$\int_{s'=0}^{S_d}\left\{\frac{q^+}{q^-}\mathcal{G}^-(s,s')\left[\frac{\Phi(s')}{2}+\phi^i(s') + \int_{s''=0}^{S_d}\mathcal{K}(s',s'')\Phi(s'')ds''\right]\right.$$
$$\left. + \mathcal{N}^-(s,s')\left[\psi^i(s') + \int_{s''=0}^{S_d}\mathcal{G}^+(s',s'')\Phi(s'')ds''\right]\right\}ds' + \frac{\psi^i(s)}{2} + \frac{1}{2}\int_{s'=0}^{S_d}\mathcal{G}^+(s,s')\Phi(s')ds' = 0, \qquad (4.202)$$

which yields, after simplification:

$$\left[\left(\mathcal{N}^- + \frac{\mathbb{I}}{2}\right)\mathcal{G}^+ + \frac{q^+}{q^-}\mathcal{G}^-\left(\mathcal{K} + \frac{\mathbb{I}}{2}\right)\right]\Phi = -\left(\mathcal{N}^- + \frac{\mathbb{I}}{2}\right)\psi^i(s) + \frac{q^+}{q^-}\mathcal{G}^-\phi^i, \qquad (4.203)$$

with the symbol $\mathcal{O}\eta$ denoting the function $\int_{s'=0}^{s_d} \mathcal{O}(s,s')\eta(s')ds'$ in operator notation.

For $z \to +\infty$, the expression of $\tilde{\Gamma}^+ \equiv F^+$ given by eq. (4.193) can be simplified since $\text{sgn}(z-z') = \pm 1$ and $|z-z'| = \pm(z-z')$, in such a way that the expression of $F^+$ at infinity becomes a sum of plane waves:

$$F^+(x,y) = \sum_{m=-\infty}^{\infty} r_m \exp\left(i\alpha_m x + i\gamma_m^+ z\right) \quad \text{if } z > z_0, \tag{4.204}$$

with amplitudes given by:

$$r_m = \frac{1}{2id\gamma_m^+} \int_{s=0}^{s_d} e^{-imKx(s) - i\gamma_m^+ z(s)} \Phi(s)\, ds. \tag{4.205}$$

The diffraction efficiencies of the reflected waves are then deduced by:

$$\rho_m = \frac{\gamma_m^+}{\gamma_0^+} |r_m|^2. \tag{4.206}$$

Similarly, it can be derived from eq. (4.201) that the transmitted field can be represented by a sum of plane waves below the profile:

$$F^-(x,y) = \sum_{m=-\infty}^{\infty} t_m \exp\left(i\alpha_m x + i\gamma_m^- z\right) \quad \text{if } z < 0. \tag{4.207}$$

The amplitudes of the transmitted plane waves are derived from eq. (4.185) after calculating the functions $\phi^-$ and $\psi^-$ from eqs. (4.199) and (4.200):

$$t_m = \frac{1}{2d} \int_{s=0}^{s_d} e^{-imKx(s) + i\gamma_m^- z(s)} \left[\frac{i\phi^-(s)}{\gamma_m^-} + \left(\frac{dx(s)}{ds} + \frac{\alpha_m}{\gamma_m^+}\frac{dz(s)}{ds}\right)\psi^-(s)\right] ds. \tag{4.208}$$

It is interesting to notice that this method has been extended to other problems (including 3D problems of scattering) by many specialists of theoretical physics and applied mathematics [49-52].